\documentclass[final, 3p, times]{elsarticle}

\usepackage{amssymb}
\usepackage{color}
\usepackage{multirow}
\journal{Journal of Computational Physics}

\begin{document}
\newcommand{\mm}{\mathrm{Minmod}}
\newcommand{\md}{\mathrm{Median}}
\newcommand{\HALF}{\frac{1}{2}}
\newcommand{\tHALF}{\frac{3}{2}}
\newcommand{\pd}[2]{\frac{\partial #1}{\partial #2}}
\newcommand{\DS}{\displaystyle}
\renewcommand{\vec}[1]{\mathbf{#1}}
\newcommand{\tens}[1]{\mathsf{#1}}
\newcommand{\REC}{{\cal R}}
\newcommand{\Lim}{LimO3$\;$}

\newcommand{\red}[1]{\color{red} #1 \color{black}}

\begin{frontmatter}



\title{High-Order Conservative Finite Difference GLM-MHD Schemes for Cell-Centered MHD}


\author[ut]{Andrea Mignone\corref{cor1}}
\ead{mignone@ph.unito.it}

\author[ut]{Petros Tzeferacos}
\ead{petros.tzeferacos@to.infn.it}

\author[ot]{Gianluigi Bodo}
\ead{bodo@oato.inaf.it}

\cortext[cor1]{Corresponding Author}

\address[ut]{Dipartimento di Fisica Generale, Universit\'a degli studi di Torino, via Pietro Giuria 1, 10125 Torino, Italy}
\address[ot]{INAF Osservatorio Astronomico di Torino, Strada Osservatorio 20, 10025 Pino Torinese, Italy}

\begin{abstract}
We present and compare third- as well as fifth-order accurate finite 
difference schemes for the numerical solution of the compressible 
ideal MHD equations in multiple spatial dimensions.
The selected methods lean on four different reconstruction techniques 
based on recently improved versions of the weighted essentially 
non-oscillatory (WENO) schemes, monotonicity preserving 
(MP) schemes as well as slope-limited polynomial reconstruction.
The proposed numerical methods are highly accurate in smooth 
regions of the flow, avoid loss of accuracy in proximity of smooth extrema and 
provide sharp non-oscillatory transitions at discontinuities.

We suggest a numerical formulation based on a cell-centered approach where all 
of the primary flow variables are discretized at the zone center. 
The divergence-free condition is enforced by augmenting the 
MHD equations with a generalized Lagrange multiplier yielding a 
mixed hyperbolic/parabolic correction, as in Dedner et al. 
(J. Comput. Phys. 175 (2002) 645-673).
The resulting family of schemes is robust, cost-effective and straightforward to 
implement. Compared to previous existing approaches, it completely avoids the 
CPU intensive workload associated with an elliptic divergence cleaning step and
the additional complexities required by staggered mesh algorithms.

Extensive numerical testing demonstrate the robustness and reliability of 
the proposed framework for computations involving both smooth and discontinuous 
features.
\end{abstract}

\begin{keyword}
Magnetohydrodynamics \sep 
Compressible Flow \sep 
Higher-order methods \sep
WENO schemes \sep 
Monotonicity Preserving \sep
Cell-centered methods



\end{keyword}

\end{frontmatter}


\section{Introduction}
\label{sec:intro}
%
%
%
%%%%%%%%%%%%%%%%%%%%%%%%%%%%%%%%%%%%%%%%%%%%%%%%%%%%%%%%%%%%%%%%

The development of high-order schemes has been receiving an 
increasing amount of attention from practitioners in the fields of 
fluid dynamics and, only more recently, magnetohydrodynamics (MHD).
This interest is driven by a variety of reasons, such as the 
possibility of obtaining highly accurate solutions with reduced 
computational effort as well as the need to narrow the gap between the 
smallest resolved features and the dissipative scales. 
Although several successful strategies have been developed in the 
context of the Euler equations of gasdynamics, only few of them 
have been extended to MHD.
In the present context, we focus our attention on 
high-order finite difference schemes for the solution of the 
compressible MHD equations in multiple spatial dimensions, 
\begin{equation}
\label{eq:mhd}
\begin{array}{rcl}
\DS \pd{\rho}{t} + \nabla\cdot\left(\rho\vec{v}\right) &=& 0 \,, \\ \noalign{\medskip}
\DS \pd{\vec{(\rho\vec{v})}}{t} + \nabla\cdot\left[
   \rho\vec{v}\vec{v}^T 
 -     \vec{B}\vec{B}^T 
 + \tens{I}\left(p + \frac{\vec{B}^2}{2}\right)\right] &=& 0\,,\\ \noalign{\medskip}
\DS \pd{\vec{B}}{t} - \nabla\times\left(\vec{v}\times\vec{B}\right) 
   &=& 0\,,\\ \noalign{\medskip}
\DS \pd{E}{t} + \nabla\cdot\left[
  \left(E + p + \frac{\vec{B}^2}{2}\right)\vec{v} - 
  \left(\vec{v}\cdot\vec{B}\right)\vec{B}\right]& =& 0\,,
\end{array}
\end{equation}
where $\rho$, $\vec{v}$, $\vec{B}$, $E$ and $p$ are
the fluid density, velocity vector, magnetic induction, energy
and gas pressure, respectively.
The system of equations (\ref{eq:mhd}) is complemented by the 
divergence-free constraint of the magnetic field,
\begin{equation}\label{eq:divB}
 \nabla\cdot\vec{B} = 0\,,
\end{equation}
and by an equation of state relating energy and pressures. 
For the present work we assume an ideal gas law
\begin{equation}
  E = \frac{p}{\Gamma-1} + \frac{1}{2}\left(\rho\vec{v}^2+\vec{B}^2\right)\,,
\end{equation}
where $\Gamma$ is the ratio of specific heats.

Traditional second-order schemes have been largely employed 
for the solution of Eq. (\ref{eq:mhd}) using either  
finite volume (FV, e.g., \citep{ZMC94, RJF95, C96, FKJ98, B98, 
DW98, FKJ98, Powell99,BS99}) or finite difference (FD, e.g., 
\cite{Balsara04, LdZ04, Crockett05, Torrilhon05, FHT06, AT08, LD09}) methods.
At the second-order level, the two approaches are essentially 
equivalent and popular schemes have been built on Godunov-type discretizations
based on the Total Variation Diminishing (TVD, \cite{Harten83}) property 
making use of slope-limited reconstructions. 
In spite of the excellent results produced in proximity of discontinuous 
waves where sharp non-oscillatory transitions can be obtained, TVD schemes 
still suffer from excessive unwanted numerical dissipation 
in regions of smooth flow. 
This deficiency owes to the inherent behavior of TVD methods that reduces the 
order of accuracy to first-order near local extrema (clipping) and smear 
linearly degenerate fields (such as contact waves) much more than shocks. 
Furthermore, discretization errors are mainly responsible 
for the loss of accuracy. 

Efforts to relax the TVD condition and overcome these limitations have 
been spent over the last decades towards the development of highly accurate 
schemes that retain the robustness common to second-order 
Godunov-type methods.
The original piecewise parabolic method (PPM) method by \cite{CW84}, 
for example, provides fourth-order accurate interface values in 
smooth regions (in 1D) and has been extended to MHD by \cite{DW94, DW97} and, 
more recently by \cite{GS05, GS08}.
PPM, however, still degenerates to first-order at 
smooth extrema and attempts to solve the problem have been 
recently presented in \cite{CS08} and \cite{RGK07}.
 
Based on a different approach, weighted essentially non-oscillatory 
(WENO, \cite{Shu97}) schemes have improved on their ENO predecessor 
(originally proposed by Harten et al. \cite{HEOC87}) and are now considered 
a powerful and effective tool for solving hyperbolic partial 
differential equations. 
WENO methods provide highly accurate solutions in regions of smooth flow
and non-oscillatory transitions in presence of discontinuous waves by 
combining different interpolation stencils of order $r$
into a weighted average of order $2r-1$. 
The nonlinear weights are adjusted by the local smoothness of 
the solution so that essentially zero weights are given to non smooth 
stencils while optimal weights are prescribed in smooth regions. 
WENO scheme have been formulated in the context of MHD using both 
FD \cite{JW99, BS00} and FV formulations,
\cite{TB04, Balsara04, DBFM08, BRDM09, Balsara09}.
Third- and fifth-order WENO schemes have been recently improved
in terms of reduced dissipation, better resolution properties
and faster convergence rates (see \cite{YC09} and \cite{BCCS08})
and will be considered here.

An alternative strategy is followed by the Monotonicity Preserving (MP) 
family of schemes by Suresh \& Huynh \cite{SH97} who proposed to carry the 
reconstruction step by first computing an accurate and stable interface value 
and then by imposing monotonicity- and accuracy-preserving constraints to 
limit the original value.
%Strangely and despite their effectiveness, monotonicity preserving schemes 
%did not receive a worth deal of attention as their WENO competitors.
MP schemes have been successfully merged with WENO methods by \cite{BS00}
and employed in the context of relativistic MHD by \cite{ECHO07}.

Finally, a reconstruction procedure that avoids the clipping phenomenon
has been recently discussed by \v{C}ada \& Torrilhon \cite{CT09} who devised a 
new class of nonlinear limiter functions based upon a non-polynomial 
reconstruction showing good shape-preserving properties.

It is important to point out that, for spatial accuracy higher than two, 
multidimensional FV schemes become notoriously more elaborate than 
their FD counterparts, since point values can no longer 
be interchanged with volume averages.
As a result, FV schemes generally require fully multidimensional 
reconstructions and the solution of several Riemann problems at a 
zone face providing the necessary number of quadrature points 
required by the desired level of accuracy, see, for instance, 
\cite{CS98, TT04, Balsara04}.
However, FV algorithms do have the adventage that they are better suited 
to non-uniform grids and adaptive mesh hierarchies.
High-order FV schemes have been recently ameliorated in the
work of \cite{DBFM08, BRDM09, Balsara09} 
using either ADER-WENO schemes or least-squares polynomial reconstruction.

Conversely, multidimensional FD schemes evolve the point values 
of the conserved quantities and considerably ease up the coding 
efforts by restricting the computations of flux derivatives 
to one dimensional stencils.
%
%This is the approach we intend to pursue here and
%, encouraged by the 
%results obtained in a previous work \cite{MT09}, 
%we will adopt a point-wise, cell-centered formulation for the magnetic field as well.
%Indeed, following \cite{Dedner02}, we enforce the divergence-free constraint 
%the induction equation is coupled to a generalized Lagrange multiplier (GLM).
%
%The resulting constrained GLM-MHD equations are still hyperbolic and fully 
%conservative in all flow variables with the exception multiplier that satisfies 
%a non-homogeneous equation.
%This amounts to a mixed hyperbolic/parabolic correction whereby 
%divergence errors are transported at the maximal admissible characteristic speed
%(regardless of the fluid velocity) and damped at the same time.
%This avoids the computational cost associated with an elliptic cleaning step
%due to a Helmholtz-Hodge projection as in \cite{JW99}, and the scrupulous 
%treatment of staggered fields demanded by constrained transport algorithms, 
%e.g. \cite{Balsara04, GS05, Balsara09}.
%
In this perspective, we present a new class of FD numerical schemes
adopting a point-wise, cell-centered formulation of all of the
flow quantities, including magnetic fields.
The proposed schemes have order of accuracy three and five and
their performance is compared through extensive testing on two 
and three-dimensional problems.
Selected third-order accurate schemes are i) an improved version of the 
classical third-order WENO scheme of \cite{JS96} based on new weight 
functions designed to improve accuracy near critical points \cite{YC09} and 
ii) the recently proposed non-polynomial reconstruction of \cite{CT09}. 
Selected fifth-order schemes include i) the WENO-Z scheme of
\cite{BCCS08} and ii) the monotonicity preserving scheme of \cite{SH97} 
based on a fifth-order accurate interface value (MP5 henceforth). 

The solenoidal constraint of the magnetic field is controlled by extending
the hyperbolic/parabolic divergence cleaning technique of Dedner et al. 
\cite{Dedner02} to FD schemes.
This avoids the computational cost associated with an elliptic cleaning step
as in \cite{JW99}, and the scrupulous treatment of staggered fields demanded 
by constrained transport algorithms, e.g. \cite{Balsara04, LdZ04, GS05, Balsara09}.
Furthermore, Mignone \& Tzeferacos \cite{MT10} have shown through extensive testing,
for a class of second-order accurate schemes, that the GLM approach is robust and can 
achieve accuracy comparable to the constrained transport.
The resulting class of schemes is explicit and fully conservative in mass, 
momentum, magnetic induction and energy.
Besides the ease of implementation and efficiency issues, the benefits offered 
by a method where all of the primary flow variables are placed at the same spatial 
position ease the task to add more complex physics.

The comparison between the different methods of solution is conveniently handled 
using the PLUTO code for computational astrophysics \cite{Mignone07}.

The paper is structured as follows. In \S\ref{sec:eqns} we describe the 
GLM-MHD equations, while \S\ref{sec:numscheme} shows the finite
difference formulation and the selected reconstruction methods.
In \S\ref{sec:numtest} we test and compare the different scheme
performance on problems involving the propagation of both 
continuous and discontinuous features.
Conclusions are drawn in \S\ref{sec:conclusions}.

%%%%%%%%%%%%%%%%%%%%%%%%%%%%%%%%%%%%%%%%%%%%%%%%%%%%%%%%%%%%%%%%%
\section{The Constrained GLM-MHD Equations}
\label{sec:eqns}
%
%
%
%%%%%%%%%%%%%%%%%%%%%%%%%%%%%%%%%%%%%%%%%%%%%%%%%%%%%%%%%%%%%%%%

We look at a conservative discretization of the MHD equations 
(\ref{eq:mhd}) where all fluid variables retain a cell-centered 
collocation and enforce the divergence-free condition through 
the hyperbolic/parabolic divergence cleaning technique of 
Dedner's \cite{Dedner02}. In this approach Gauss's and Faraday's laws 
of magnetism are modified by the introduction of a new scalar 
field function or generalized Lagrangian multiplier 
(GLM henceforth) $\psi$.
The resulting system of GLM-MHD equations then reads
\begin{equation}\label{eq:glm_mhd}
\pd{\vec{U}}{t} = -\sum_{l=x,y,z} \pd{\vec{F}_l}{l} + \vec{S}\,,
\end{equation}
with conservative state vector $\vec{U}$ and fluxes $\vec{F}_l$ 
defined by
\begin{equation}\label{eq:state_and_flux}
\vec{U}= \left(\begin{array}{c}
 \rho     \\ \noalign{\medskip}
 \rho v_d \\ \noalign{\medskip}
 B_d \\ \noalign{\medskip}
 E        \\ \noalign{\medskip}
 \psi \end{array}\right)
\,,\quad
\vec{F}_l = \left(\begin{array}{c}
 \rho v_l  \\ \noalign{\medskip}
 \rho v_dv_l - B_dB_l + \delta_{dl}\left(p + \vec{B}^2/2\right)  \\ \noalign{\medskip}
  B_dv_l - B_lv_d + \delta_{dl}\psi \\ \noalign{\medskip}
 \left(E + p + \vec{B}^2/2\right)v_l - \left(\vec{v}\cdot\vec{B}\right)B_l    \\ \noalign{\medskip}
 c_h^2 B_l \end{array}\right)
\,,\quad
 \vec{S} = 
\left(\begin{array}{c}
  0 \\ \noalign{\medskip}
  0 \\ \noalign{\medskip}
  0 \\ \noalign{\medskip}
  0 \\ \noalign{\medskip}
 -c_h^2/c_p^2\psi  \end{array}\right) \,,
\end{equation}
where $d = x,y,z$ labels the different components while $\delta_{dl}$ is 
the delta Kronecker symbol.
Equations (\ref{eq:glm_mhd}) are hyperbolic and fully conservative 
with the only exception of the unphysical scalar field $\psi$ which 
satisfies a non-homogeneous equation with a source term. 
In the GLM approach, divergence errors are propagated to the domain boundaries at 
finite speed $c_h$ and damped at a rate given by $c_h^2/c_p^2$
(see \S\ref{sec:eqns}).

The eigenvalues of the MHD flux Jacobians $\partial\vec{F}_l/\partial\vec{U}$ 
are all real and coincide with the ordinary MHD waves plus two additional 
modes $\pm c_h$, for a total of $9$ characteristic waves.
Restricting our attention to the $l=x$ direction, they are given by
\begin{equation}
 \lambda^{1,9} = \mp c_h     \,,\quad
 \lambda^{2,8} = v_x \mp c_f \,,\quad
 \lambda^{3,7} = v_x \mp c_a \,,\quad
 \lambda^{4,6} = v_x \mp c_s \,,\quad
 \lambda^5     = v_x \,,
\end{equation}
where
\begin{equation}\label{eq:eigenvalues}
c_{f,s} = \sqrt{\frac{1}{2\rho}\left(\Gamma p + |\vec{B}|^2 \pm 
          \sqrt{\left(\Gamma p + |\vec{B}|^2\right)^2 
                - 4\Gamma p B_x^2}\,\right)}
\,,\quad
c_a = \frac{\left|B_x\right|}{\sqrt{\rho}} \,,
\end{equation}
are the fast magneto-sonic ($c_f$ with the $+$ sign), 
slow magneto-sonic ($c_s$ with the $-$ sign) and Alfv\'en velocities.       
The two additional modes $\pm c_h$ are decoupled from the remaining ones 
and corresponds to linear waves carrying jumps in $B_x$ and $\psi$.
These waves are made to propagate at the maximum signal speed compatible
with the time step, i.e., 
\begin{equation}\label{eq:ch}
 c_h = \max\left(|v_x| + c_{f,x}, |v_y| + c_{f,y}, |v_z| + c_{f,z}\right)\,.
\end{equation}
where $c_{f,x}, c_{f,y}, c_{f,z}$ are the fast magneto-sonic speeds in the 
three directions and the maximum is taken throughout the domain.

Owing to the decoupling, one can treat the $2\times 2$ linear
system given by the longitudinal component of the field $B_l$ and 
$\psi$ separately from the other ordinary $7$-wave MHD equations.
As we shall see, this greatly simplifies the solution process and 
allows to use the standard characteristic decomposition of the MHD 
equations.

%The system of equations given in (\ref{eq:glm_mhd}) is advanced in time
%by solving the homogeneous part separately from the source term 
%contribution, in an operator-split fashion.
%These are the subjects of the next section.

Following \cite{MT10}, we divide the solution process into an homogeneous step, 
where the GLM-MHD (\ref{eq:glm_mhd}) are solved with $\vec{S}=0$, and a source step, 
where integration is done analytically:
\begin{equation} \label{eq:source_exp}
 \psi^{(\Delta t)} = \psi^{(0)}
  \exp\left(-\alpha_p\frac{c_h}{\Delta h/\Delta t}\right)\,,
 \quad \mathrm{with} \quad
  \alpha_p = \Delta h\,\frac{c_h}{c_p^2}\,.
\end{equation}
where $\Delta h = \min(\Delta x, \Delta y, \Delta z)$ is the minimum 
grid size.
Extensive numerical testing has shown that divergence errors are minimized when
the parameter $\alpha_p$ lies in the range $[0,1]$ depending on the particular problem, 
although in presence of smooth flows this choice seems to be less sensitive to the 
numerical value of $\alpha_p$.

%%%%%%%%%%%%%%%%%%%%%%%%%%%%%%%%%%%%%%%%%%%%%%%%%%%%%%
\section{Finite Difference schemes}
\label{sec:numscheme}
%
%
%
%
%
%%%%%%%%%%%%%%%%%%%%%%%%%%%%%%%%%%%%%%%%%%%%%%%%%%%%%%

We consider a conservative finite difference discretization 
of (\ref{eq:glm_mhd}) where point-values rather than volume 
averages are evolved in time.
A uniform Cartesian mesh is employed with cell sizes 
$\Delta x\times\Delta y\times\Delta z$ centered
at $(x_i,y_j,z_k)$, where $i,j,k$ label the computational 
zones in the three directions.
For clarity of exposition, we disregard the integer subscripts
when redundant but always keep the half increment index notation when 
referring to a cell boundary, e.g., 
$\vec{F}_{i+\HALF} \equiv \vec{F}_{i+\HALF,j,k}$.

Integration in time resorts to a semi-discrete formulation where,
given a high-order numerical approximation ${\cal L}(\vec{U})$ 
to the derivatives appearing on the right hand side of 
Eq. (\ref{eq:glm_mhd}), one is faced with the solution of the 
following initial value problem 
\begin{equation}\label{eq:ivp}
 \frac{d\vec{U}}{dt} = \vec{\cal{L}}\left(\vec{U}\right) \,,
\end{equation}
with initial condition given by the point-wise values
of $\vec{U}(x_i,y_j,z_k,t^n) \equiv \vec{U}^n_{i,j,k}$.
We choose the popular  third-order Runge-Kutta 
scheme \cite{Shu88,GS98} to advance the solution in time, for which one has
\begin{equation}\label{eq:rk3}
 \begin{array}{lcl}
 \vec{U}^*     &=& \vec{U}^n + {\cal L}\left(\vec{U}^n\right) \,,\\ \noalign{\medskip}
 \vec{U}^{**}  &=& \DS \frac{3}{4}\vec{U}^n + \frac{1}{4}\vec{U}^* 
                     + \frac{\Delta t^n}{4}{\cal L}\left(\vec{U}^*\right)\,, \\ \noalign{\medskip}
 \vec{U}^{n+1} &=& \DS \frac{1}{3}\vec{U}^n + \frac{2}{3}\vec{U}^{**} 
                     + \frac{2}{3}\Delta t^n{\cal L}\left(\vec{U}^{**}\right)\,.
\end{array}
\end{equation}
The choice of the time step $\Delta t^n$ is restricted by
the Courant-Friedrichs-Levy (CFL) condition:
\begin{equation}\label{eq:deltat}
 \Delta t^n = C_a\frac{\Delta h}{c_h}\,,
\end{equation}
where $C_a$ is the CFL number.
Since the time step is proportional to the mesh size, the overall accuracy 
of the scheme is restricted to third-order because of the time-stepping 
introduced in Eq. (\ref{eq:rk3}).

Our task is now to provide a stable and accurate non-oscillatory
numerical approximation to ${\cal L}(\vec{U})$.
To this purpose, we begin by focusing our attention to the $x-$ direction 
and set, for ease of notations, $\vec{F}_i\equiv \vec{F}_x(\vec{U}_{i,j,k})$. 
We then let point values of the flux $\vec{F}_i$ correspond to the 
volume averages of another function, say $\hat{\vec{F}}$, and define
\begin{equation}\label{eq:flux_primitive}
  \vec{F}_i = \frac{1}{\Delta x}\int_{x_{i-\HALF}}^{x_{i+\HALF}}  \hat{\vec{F}}(\xi)d\xi
%    =\frac{1}{\Delta x}\left[\int_{-\infty}^{x_{i+\HALF}} \hat{\vec{F}}(\xi)d\xi- 
%                             \int_{-\infty}^{x_{i-\HALF}} \hat{\vec{F}}(\xi)d\xi\right]
    = \frac{1}{\Delta x}\Big[\vec{H}(x_{i+\HALF}) - \vec{H}(x_{i-\HALF})\Big]
\,,\quad\textrm{where}\quad
\vec{H}(x) = \int_{-\infty}^x \hat{\vec{F}}(\xi)d\xi\,.
\end{equation}
In this formalism, point values of the flux $\vec{F}_i$ are identified as 
cell averages of $\hat{\vec{F}}(x)$ and $\vec{H}(x)$ may be regarded as 
the primitive function of $\hat{\vec{F}}$.
Straightforward differentiation of Eq. (\ref{eq:flux_primitive}) yields
the conservative approximation
\begin{equation}\label{eq:dfdx}
 \left.\pd{\vec{F}}{x}\right|_{x_i} = \frac{1}{\Delta x}
  \left(\hat{\vec{F}}_{i+\HALF} - \hat{\vec{F}}_{i-\HALF}\right)\,.
\end{equation}
Stated in this form, the problem consists of finding a high-order approximation 
to the interface values of $\hat{\vec{F}}_{i+\HALF}$ knowing the 
undivided differences of the primitive function 
$\vec{H}(x)$, a procedure entirely analogous to that 
used in the context of finite volume methods such as PPM \cite{CW84}.
Thus one can set
\begin{equation}
 \hat{\vec{F}}_{i+\HALF} = \REC\left(\vec{F}_{[s]}\right)\,,
\end{equation}
where $\REC()$ is a highly accurate reconstruction scheme
providing a stable interface flux value from point-wise values 
%and $2S+2$ defines the interpolation stencil. 
and the index $[s]$ spans through the interpolation stencil.

The procedure can be repeated in an entirely similar way also for the 
$y$ and $z$ flux contributions and allows to write the ${\cal L}$ operator
in (\ref{eq:ivp}) as
\begin{equation}\label{eq:L}
 {\cal L}\left(\vec{U}\right) = 
  -\frac{1}{\Delta x}\left(\hat{\vec{F}}_{x,i+\HALF} - \hat{\vec{F}}_{x,i-\HALF}\right)
  -\frac{1}{\Delta y}\left(\hat{\vec{F}}_{y,j+\HALF} - \hat{\vec{F}}_{y,j-\HALF}\right)
  -\frac{1}{\Delta z}\left(\hat{\vec{F}}_{z,k+\HALF} - \hat{\vec{F}}_{z,k-\HALF}\right)\,.
\end{equation}

This yields the fully unsplit approach considered in this paper.
Alternatively, one could use a directionally split formalism to 
obtain the solution through a sequence of one dimensional problems 
separately corresponding to each term in equation (\ref{eq:L}).

%Note that we do not employ directional splitting to evolve 
%the equations in time but include the contribution
%from all fluxes simultaneously.
%\red{Although this choice gives
%a more severe restriction on the time step, it was nevertheless
%shown to yield smaller divergence errors.}

In order to ensure robustness and to avoid the appearance of spurious 
oscillations, the reconstruction step is best carried with the help 
of local characteristic fields and by separately evaluating contributions 
coming from right- and left-going waves.
To this end we first compute, using the simple
arithmetic average $\vec{U}_{i+\HALF} = (\vec{U}_i + \vec{U}_{i+1})/2$, 
left and right eigenvectors $\vec{L}^{\kappa}_{i+\HALF}$ and 
$\vec{R}^{\kappa}_{i+\HALF}$ of the Jacobian matrix 
$\partial\vec{F}/\partial\vec{U}$, for each characteristic field $\kappa=1,\dots,9$.
We then obtain a projection of the positive and negative part of the flux 
using a simple Rusanov Lax-Friedrichs flux splitting:
\begin{equation}\label{eq:char_vars}
 \left\{\begin{array}{rcl}
 V^{\kappa,+}_{i+\HALF,[s]} &=& \frac{1}{2}\vec{L}^{\kappa}_{i+\HALF}\cdot
                                \left(\vec{F}_{[s]} + \alpha^{\kappa}\vec{U}_{[s]}\right) \,,
   \\ \noalign{\medskip}
 V^{\kappa,-}_{i+\HALF,[s]} &=& \frac{1}{2}\vec{L}^{\kappa}_{i+\HALF}\cdot
                                \left(\vec{F}_{[s']} - \alpha^{\kappa}\vec{U}_{[s']}\right) \,,
 \end{array}\right.   
\end{equation}
where $\vec{F}_{[s]}$ and $\vec{U}_{[s]}$ are the point-wise values of the flux 
and conservative variables.
For a typical one-point upwind-biased approximation of order $(2r+1)$, one has 
$[s] = i-r,\dots,i+r$ while $[s'] = 2i-[s]+1$ mirrors left-going characteristic 
fields with respect to the interface $i+\HALF$.
The coefficient $\alpha^{\kappa}$ represents the maximum absolute value 
of the $\kappa$-th characteristic speed throughout the domain.

The global Lax-Friedrichs flux splitting thus introduced
is particularly diffusive and other forms of splitting are of course possible,
e.g. \cite{JS96,BS00}.
However, we have found that the level of extra numerical dissipation tend to 
become less important for higher-order scheme.

The interface flux is then written as a local expansion in the 
right-eigenvector space:
\begin{equation}\label{eq:flux_pm}
 \hat{\vec{F}}_{i+\HALF} = \sum_{\kappa} 
 \left(\hat{V}^{\kappa,+}_{i+\HALF} 
     + \hat{V}^{\kappa,-}_{i+\HALF}\right)\vec{R}^{\kappa}_{i+\HALF} 
\,,
\end{equation}
where the coefficients
\begin{equation}\label{eq:interp}
 \hat{V}^{\kappa,\pm}_{i+\HALF} \equiv \REC\left(V^{\kappa,\pm}_{i+\HALF,[s]}\right)\,.
\end{equation}
are the reconstructed interface values of the local characteristic fields
and $\REC()$ can be any one of the procedures described in \S\ref{sec:recon}.
%In the following sections we will describe different choices for $\REC$ and
%set, for simplicity, $f_s= V^{\kappa,\pm}_{[i+\HALF],s}$

%%%%%%%%%%%%%%%%%%%%%%%%%%%%%%%%%%%%%%%%%%%%%%%%%%%%%%%%%%%%%%%%%%%%%
\subsection{Modification for the Constrained GLM-MHD equations}
\label{sec:glm_mod}
%
%
%%%%%%%%%%%%%%%%%%%%%%%%%%%%%%%%%%%%%%%%%%%%%%%%%%%%%%%%%%%%%%%%%%%%%

The procedure illustrated so far is valid for an arbitrary system of 
hyperbolic conservation laws, provided $\vec{L}^{\kappa}$ and $\vec{R}^{\kappa}$ 
satisfy
\begin{equation}
 \vec{L}^{\kappa}\cdot\pd{\vec{F}}{\vec{U}}\cdot\vec{R}^{\kappa} = \lambda^{\kappa}\,,
\end{equation}
i.e., they are left and right eigenvectors of the flux Jacobian, respectively.
However, following \cite{Dedner02}, we wish to exploit the full 
$7\times 7$ characteristic decomposition of the usual MHD equations
rather than resorting to a full $9\times 9$ diagonalization procedure.
To this purpose, we take advantage of the fact that the longitudinal component 
of the field $B_x$ and the Lagrange multiplier $\psi$ satisfy 
\begin{equation}\label{eq:2x2}
 \pd{}{t}\left(\begin{array}{c}
 B_x \\ \noalign{\medskip}
 \psi 
 \end{array}\right) 
 + 
 \left(\begin{array}{cc}
 0     & 1 \\ \noalign{\medskip}
 c_h^2 & 0 
 \end{array}\right) \pd{}{x}\left(\begin{array}{c}
  B_x \\ \noalign{\medskip}
  \psi 
 \end{array}\right) = 0 \,,
\end{equation}
and are thus decoupled from the remaining seven MHD equations.
Eq. (\ref{eq:2x2}) defines a constant coefficient
linear hyperbolic system with left and right 
eigenvectors given, respectively, by the rows and columns of 
\begin{equation}\label{eq:2x2_evect}
\tens{L}_{2\times 2} = \frac{1}{2}\left(\begin{array}{cc}
   1    &  -1/c_h   \\ \noalign{\medskip}
   1    &   1/c_h 
\end{array}\right) 
\,,\qquad
 \tens{R}_{2\times 2} = \left(\begin{array}{cc}
   1    &  1   \\ \noalign{\medskip}
  -c_h  & c_h 
\end{array}\right) \,,
\end{equation}
associated with the eigenvalues $\lambda^1=-c_h$ and $\lambda^9=+c_h$.
The $2\times 2$ linear system (\ref{eq:2x2}) can be preliminary solved 
to find the values of $B_x$ and $\psi$ at a given interface.
Indeed, by applying the projection (\ref{eq:char_vars}) to the
linear system (\ref{eq:2x2}) using Eq. (\ref{eq:2x2_evect}), 
one obtains that the only non trivial characteristic fields are
\begin{equation}\label{eq:k19}
  V^{1,-}_{i+\HALF,[s]} = \frac{1}{2}\left(\psi_{[s']} - c_hB_{x,[s']}\right)\,,\quad
  V^{9,+}_{i+\HALF,[s]} = \frac{1}{2}\left(\psi_{[s]}  + c_hB_{x,[s]}\right)\,.
\end{equation}
Since the eigenvectors are constant in space, the local projection at 
$i+\HALF$ are completely unnecessary and the computations in Eq. 
(\ref{eq:k19}) can be carried out very efficiently throughout the grid.
Once (\ref{eq:k19}) have been reconstructed using Eq. (\ref{eq:interp}) 
one defines
\begin{equation}\label{eq:bx_intrfc}
 B_{x,i+\HALF}   = \left(\hat{V}^{9,+}_{i+\HALF} - \hat{V}^{1,-}_{i+\HALF}\right)/c_h \,,\quad
\psi_{x,i+\HALF} = \hat{V}^{9,+}_{i+\HALF} + \hat{V}^{1,-}_{i+\HALF}\,,
\end{equation}
and proceed by solving the ordinary $7\times 7$ MHD equations 
using $B_{x,i+\HALF}$ defined by (\ref{eq:bx_intrfc}) as a constant 
parameter.

%%%%%%%%%%%%%%%%%%%%%%%%%%%%%%%%%%%%%%%%%%%%%%%%%%%%%%%%%%%%%%%%%%%%%%%%
\subsection{Third and Fifth-order Accurate Reconstructions}
\label{sec:recon}
%
%
%
%%%%%%%%%%%%%%%%%%%%%%%%%%%%%%%%%%%%%%%%%%%%%%%%%%%%%%%%%%%%%%%%%%%%%%%%

%In \S\ref{sec:numscheme} it was shown that the spatial derivatives of the 
%flux, Eq. (\ref{eq:dfdx}), are conveniently cast in conservative 
%form by providing accurate interface values through a finite volume 
%reconstruction approach.
%Indeed, in virtue of Eq (\ref{eq:flux_primitive}), one could interpolate
%the point-wise flux values $\vec{F}_i$ to the cell interface using any 
%standard finite volume algorithm.

We have shown in \S\ref{sec:numscheme} that flux derivatives may
be written in conservative form by applying any one-dimensional finite 
volume reconstruction to the point values of the flux 
$\vec{F}_i$.
Among the variety of different strategies we investigate both third-
and fifth-order accurate interpolation schemes making use of 
three- and five-point stencil, respectively:
\begin{itemize}
 \item an improved version of the classical third-order WENO scheme
       of \cite{JS96} based on new weight functions designed to improve accuracy 
       near critical points (WENO+3, \S\ref{sec:weno3});
 \item the recently proposed \Lim third-order reconstruction of \cite{CT09}, 
       \S\ref{sec:limo3}. 
 \item the improved WENO5 scheme of \cite{BCCS08} also known as 
       WENO-Z (\S\ref{sec:wenoz});
 \item the monotonicity preserving scheme of \cite{SH97} based on a 
       fifth-order interface value (MP5, \S\ref{sec:mp5}).
\end{itemize}
Our choice is motivated by the sake of comparing 
well-known and recently presented state of the art algorithms
that rely on heavy usage of conditional statements (\Lim and MP5)
or completely avoid them (WENO+3 and WENO-Z).

%In the present context we wish to consider four different schemes based on
%third as well as fifth-order accurate interpolation methods.
%Third-order accurate schemes make use of a three-point local stencil to 
%reconstruct each characteristic field to the zone boundary.
%Two different schemes are hereby considered:
%i) an improved version of the classical third-order WENO scheme
%of \cite{JS96} based on new weight functions designed to improve accuracy 
%near critical points (\S\ref{sec:weno3}) and ii) the recently proposed 
%non polynomial reconstruction of \cite{CT09}, described in \S\ref{sec:limo3}. 
%On the other hand, fifth-order schemes employs a five-point 
%stencil to reach the desired accuracy and are thus more expensive. 
%We have chosen two popular schemes: i) the improved WENO5 scheme of
%\cite{BCCS08} also known as WENO-Z and described in 
%\S\ref{sec:wenoz}, and ii) the monotonicity 
%preserving scheme of \cite{SH97} based on a fifth-order interface value
%(MP5 henceforth), described in \S\ref{sec:mp5}.

%Although reconstruction algorithms may be applied to the 
%point values of the flux $\vec{F}_i$ in a cheap component-wise manner, 
%we prefer to adopt the characteristic decomposition for a better-behaved 
%scheme.
The proposed algorithms are applied to the left (-) and right (+)
propagating characteristic fields defined by Eq. 
(\ref{eq:char_vars}) to provide an accurate interface
value, formally represented by Eq. (\ref{eq:interp}).
Thus, in our formulation, the total number of reconstruction is
$16$: two for the linear characteristic fields defined by 
Eq. (\ref{eq:k19}) and $14$ for the left- and right-going 
wave families defined by Eq. (\ref{eq:char_vars}) with $k=2,\dots,8$.

In the following we will drop the $i+\HALF$ index for the sake of exposition
and shorten either one of (\ref{eq:char_vars}) with $f_{[s]}$.
Undivided difference will be frequently used and denoted with
\begin{equation}
 \Delta_{i+\HALF} = f_{i+1}-f_i \,.
\end{equation}
Occasionally, we will also make use of the $\mm$ and $\md$ functions defined, 
respectively as
\begin{equation}
\mm(a,b) = \frac{\mathrm{sgn}(a)+\mathrm{sgn}(b)}{2}\min\left(|a|,|b|\right)
\,,\quad
 \md(a,b,c) = a + \mm(b-a,c-a)\,.
\end{equation}

%%%%%%%%%%%%%%%%%%%%%%%%%%%%%%%%%%%%%%%%%%%%%%%%%%%%%%%%%%%%%%%%%%%%%%%%
\subsubsection{Third-Order Improved WENO (WENO+3)}\label{sec:weno3}
%
%
%
%%%%%%%%%%%%%%%%%%%%%%%%%%%%%%%%%%%%%%%%%%%%%%%%%%%%%%%%%%%%%%%%%%%%%%%%

In the classical third-order WENO scheme of \cite{JS96}, 
the interface value is reconstructed using the information available 
on a three-point local stencil $(x_{i-1},x_i,x_{i+1})$.
More specifically, a third-order accurate value is provided by a linear 
convex combination of second-order fluxes:
\begin{equation}\label{eq:w3}
  \REC\left(f_{[s]}\right) =   \omega_0\frac{ f_i + f_{i+1}}{2} 
                             + \omega_1\frac{-f_{i-1} + 3f_{i}}{2}\,.
\end{equation}
The weights $\omega_l$ for $l=0,1$ are defined by 
\begin{equation}\label{eq:w3:omega}
  \omega_l = \frac{\alpha_l}{\sum_m\alpha_m} \,,\quad
  \alpha_l = \frac{d_l}{(\beta_l+\epsilon)^2} \,,\quad\mathrm{with}\quad
 \beta_0 = \DS \Delta_{i+\HALF}^2  \,,\quad
 \beta_1 = \DS \Delta_{i-\HALF}^2  \,,
\end{equation}
where $d_0 = 2/3, d_1=1/3$ are optimal weights and 
the smoothness indicators $\beta_l$ give a measure of 
the regularity of the corresponding polynomial approximation.

The scheme has been recently improved in the work by Yamaleev \& Carpenter, 
\cite{YC09}, where the introduction of an additional nonlinear artificial 
dissipation term was shown to make the scheme stable in the L2-energy norm 
for both continuous and discontinuous solutions.
Yamaleev \& Carpenter also derived new weight functions providing 
faster convergence and improved accuracy at critical points. 
The improved weights are still defined by Eq. (\ref{eq:w3:omega}) with 
$\alpha_l$ replaced by 
\begin{equation}\label{eq:w3:newomega}
  \alpha_l \to d_l\left(1 + \frac{\left|\Delta_{i+\HALF}-\Delta_{i-\HALF}\right|^2}
   {\beta_l+\epsilon}\right) \,.
\end{equation}
To avoid loss of accuracy at critical points, it was shown in \cite{YC09}
that $\epsilon$ has to satisfy $\epsilon = O(\Delta x^2)$. 

Here adopt the conventional third-order scheme defined by 
Eq. (\ref{eq:w3})-(\ref{eq:w3:omega}) but with $\alpha_l$ replaced
by Eq. (\ref{eq:w3:newomega}) and simply set $\epsilon = \Delta x^2$.
This improves the accuracy over the original $3^{\rm rd}$ order scheme of 
\cite{JS96} in regions where the solution is smooth and provides essentially 
non-oscillatory solutions near strong discontinuities and unresolved features.
The improved third-order WENO scheme just described will be referred to as 
WENO+3.

%%%%%%%%%%%%%%%%%%%%%%%%%%%%%%%%%%%%%%%%%%%%%%%%%%%%%%%%%%%%%%%%%%%%%%%%
\subsubsection{Third-Order Limited reconstruction (\Lim)}
\label{sec:limo3}
%
%
%
%%%%%%%%%%%%%%%%%%%%%%%%%%%%%%%%%%%%%%%%%%%%%%%%%%%%%%%%%%%%%%%%%%%%%%%%

Recently, \v{C}ada and Torrilhon \cite{CT09} have proposed a new and efficient third-order 
limiter function in the context of finite volume schemes. 
Similarly to the $3^{\rm rd}$-order WENO scheme described in \S\ref{sec:weno3}, the new 
limiter employs a local three-point stencil to achieve piecewise-parabolic reconstruction 
for smooth data and preserves the accuracy at local extrema, thus avoiding the well 
known clipping of classical second-order TVD limiters.
Interface values are reconstructed using a simple piecewise-linear max/min function
acting as a logical switch depending on the left and right slope:
\begin{equation}\label{eq:LimO3}
%\REC\left(f_s\right) = f_i + \frac{s_+}{2}{\cal L}^3\left(\Delta_{i+\HALF},\Delta_{i-\HALF}\right)
\REC\left(f_{[s]}\right) = f_i + \frac{\Delta_{i+\HALF}}{2}
  \left[P_3(\theta) + \chi\left(\hat{\phi}(\theta)-P_3(\theta)\right)\right] \,,
\end{equation}
where $\theta = \Delta_{i-\HALF}/\Delta_{i+\HALF}$ is the slope ratio, 
$P_3(\theta) = (2+\theta)/3$ is the building block giving polynomial 
quadratic reconstruction and $\hat{\phi}(\theta)$ is the third-order limiter
\begin{equation}\label{eq:lim3}
\hat{\phi}(\theta) = \left\{\begin{array}{ll}
\DS \max\left[0, \min\left(P_3(\theta),2\theta,1.6\right)\right] & \quad\mathrm{if}\quad \theta \ge 0 \,,
  \\ \noalign{\medskip}
\DS\max\left[0, \min\left(P_3(\theta),-\frac{\theta}{2}\right)\right] & \quad\mathrm{if}\quad \theta < 0 \,.
\end{array}\right.
\end{equation}

The function $\chi$ in Eq. (\ref{eq:LimO3}) smoothly switches 
between limited and unlimited reconstructions based on a local 
indicator function $\eta$ properly introduced to avoid loss
of accuracy at smooth extrema with one vanishing lateral derivative:
\begin{equation}\label{eq:alphaeta}
 \chi = \max\left[0, \min\left(1,\frac{1}{2}+\frac{\eta-1}{2\epsilon}\right)\right] \,,\qquad
 \eta = \frac{\Delta_{i-\HALF}^2 + \Delta_{i+\HALF}^2}{(r\Delta x)^2}\,,
\end{equation}
where $\epsilon = 10^{-12}$.
The function $\eta$ measures the curvature of non-monotone data
inside a computational zone and the free-parameter $0\le r\le 1$ is used to 
discriminate between smooth extrema and shallow gradients. 
Larger values of $r$ noticeably improve the reconstruction properties 
at the cost of introducing more local variation, see \cite{CT09}.
In the tests presented here we use $r=1$.

%%%%%%%%%%%%%%%%%%%%%%%%%%%%%%%%%%%%%%%%%%%%%%%%%%%%%%%%%%%%%%%%%%%%%%%%
\subsubsection{Fifth-Order Improved WENO: WENO-Z}
\label{sec:wenoz}
%
%
%
%%%%%%%%%%%%%%%%%%%%%%%%%%%%%%%%%%%%%%%%%%%%%%%%%%%%%%%%%%%%%%%%%%%%%%%%

Borges et al. \cite{BCCS08} presented an improved version of the 
classical fifth-order weighted essentially non-oscillatory (WENO) 
FD scheme of \cite{JS96}. 
The new scheme, denoted with WENO-Z, has been shown to be less dissipative and 
provide better resolution at critical 
points at a very modest additional computational cost.
We will employ such scheme here and, for the sake of completeness, 
report only the essential steps for its implementation
(for a thorough discussion see the paper by \cite{BCCS08}).

Following the general idea of WENO reconstruction, one considers
the convex combination of different third-order accurate interface 
values built on the three possible sub-stencils of $i-2\le s\le i+2$:
\begin{equation}
  \REC\left(f_{[s]}\right) =   
   \omega_0\frac{ 2f_{i-2} - 7f_{i-1}  + 11f_{i}}{6} 
 + \omega_1\frac{- f_{i-1} + 5f_{i}   +  2f_{i+1}}{6}
 + \omega_2\frac{ 2f_{i}   + 5f_{i+1} -   f_{i+2}}{6}  \,.
\end{equation}
The weights $\omega_l$ for $l=0,1,2$ are defined by 
\begin{equation}
  \omega_l = \frac{\alpha_l}{\sum_m\alpha_m} \,,\quad
  \alpha_l = \left\{\begin{array}{ll}
 \DS \frac{d_l}{(\beta_l+\epsilon)^2}  & \quad\mathrm{(WENO5)} \\ \noalign{\medskip}
 \DS d_l\left(1 + \frac{|\beta_0-\beta_2|}{\beta_l+\epsilon}\right) & \quad\mathrm{(WENO-Z)}
\end{array}\right.
\end{equation}
where $d_0 = 1/10, d_1=3/5, d_2=3/10$ are the optimal weights giving 
a fifth-order accurate approximation, $\epsilon = 10^{-40}$ is a 
small number preventing division by zero and the smoothness indicators 
$\beta_l$ give a measure of the regularity of the corresponding 
polynomial approximation:
\begin{equation}\label{eq:weno-zbeta}\begin{array}{rcl}
 \beta_0 &=& \DS \frac{13}{12}\left(\Delta_{i-\HALF} - \Delta_{i-\tHALF}\right)^2 + 
                  \frac{1}{4}\left(3\Delta_{i-\HALF} - \Delta_{i-\tHALF}\right)^2 \,,
       \\ \noalign{\medskip}
 \beta_1 &=& \DS \frac{13}{12}\left(\Delta_{i+\HALF} - \Delta_{i-\HALF}\right)^2 + 
            \frac{1}{4}\left(\Delta_{i+\HALF} + \Delta_{i+\HALF}\right)^2 \,,
       \\ \noalign{\medskip}
 \beta_2 &=& \DS \frac{13}{12}\left(\Delta_{i+\tHALF} -  \Delta_{i+\HALF}\right)^2 + 
                 \frac{1}{4}  \left(3\Delta_{i+\HALF} - \Delta_{i+\tHALF}\right)^2 \,.
\end{array}\end{equation}
While maintaining the essentially non-oscillatory behavior, the new formulation 
makes use of higher-order information about the regularity 
of the solution thus providing enhanced order of convergence at critical points
as well as reduced dissipation at discontinuities.

%%%%%%%%%%%%%%%%%%%%%%%%%%%%%%%%%%%%%%%%%%%%%%%%%%%%%%%%%%%%%%%%%%%%%%%%
\subsubsection{Fifth-order Monotonicity Preserving (MP5)}
\label{sec:mp5}
%
%
%
%%%%%%%%%%%%%%%%%%%%%%%%%%%%%%%%%%%%%%%%%%%%%%%%%%%%%%%%%%%%%%%%%%%%%%%%

The monotonicity preserving (MP) schemes of Suresh \& Huynh \cite{SH97} 
achieve high-order interface reconstruction by first providing an accurate polynomial 
interpolation and then by limiting the resulting value so as to preserve monotonicity 
near discontinuities and accuracy in smooth regions.
The MP algorithm is better sought on stencils with five or more points
in order to distinguish between local extrema and a genuine $O(1)$ 
discontinuities.
Here we employ the fifth-order accurate scheme based on 
the (unlimited) interface value given by 
\begin{equation}\label{eq:mp_hi}
  f_{i+\HALF} = \frac{2f_{i-2}-13f_{i-1} + 47f_i + 27f_{i+1} - 3f_{i+2}}{60} \,,\quad
\end{equation}
based on the five point values $f_{i-2},\dots,f_{i+2}$.
Together with (\ref{eq:mp_hi}), we also define the monotonicity-preserving 
bound
\begin{equation}\label{eq:mp_mp}
  f^{\rm MP} = f_{i} + \mm\left(\Delta_{i+\HALF}, \alpha\Delta_{i-\HALF}\right)\,,
\end{equation}
resulting from the median between $f_i$, $f_{i+1}$ and the left-sided extrapolated upper 
limit $f^{\rm UL} = f_i + \alpha\Delta_{i-\HALF}$.
The parameter $\alpha \ge 2$ controls the maximum steepness of the left sided
slope and preserves monotonicity during a single Runge-Kutta 
stage (Eq. \ref{eq:rk3}) provided the CFL number satisfies 
$C_a \le 1/(1+\alpha)$. In practice, setting $\alpha=4$ still allows 
larger values of $C_a$ to be used.
The interface value given by Eq. (\ref{eq:mp_hi}) is not altered when the data 
is sufficiently smooth or monotone that $f_{i+\HALF}$ lies inside the interval 
defined by $[f_i,f^{\rm MP}]$.
Otherwise limiting takes place by bringing the original value back
into a new interval $I[f^{\min},f^{\max}]$ specifically designed to 
preserve accuracy near smooth extrema and provide monotone profile 
close to discontinuous data.
The final reconstruction can be written as
\begin{equation}
 \REC\left(f_{[s]}\right) = \left\{\begin{array}{ll}
 f_{i+\HALF} & \quad\mathrm{if}\quad (f_{i+\HALF}-f_i)(f_{i+\HALF} - f^{\rm MP})<0 \,,
 \\ \noalign{\medskip}
 \md\left(f^{\min}, f_{i+\HALF},f^{\max}\right) 
 & \quad\mathrm{otherwise} \,,
\end{array}\right.
\end{equation}
where 
\begin{equation}\label{eq:mp_fminmax}
\begin{array}{rcl}
 f^{\min} &=& \DS \max\left[\min\left(f_i,f_{i+1}, f^{\rm MD}\right),\,
                                \min\left(f_i,f^{\rm UL}, f^{\rm LC}\right)\right]\,,
 \\ \noalign{\medskip}
 f^{\max} &=& \DS\min\left[\max\left(f_i,f_{i+1}, f^{\rm MD}\right),\,
                                \max\left(f_i,f^{\rm UL}, f^{\rm LC}\right)\right]\,.
\end{array}
\end{equation}
The bounds given by Eq. (\ref{eq:mp_fminmax}) provide accuracy-preserving 
constraints by allowing the original interface value $f_{i+\HALF}$ to lie
in a somewhat larger interval than $I[f_i,f_{i+1}]$ or $I[f_i,f^{\rm UL}]$.
This is accomplished by considering the intersection of the two extended 
intervals $I[f_i,f_{i+1},f^{\rm MD}]$ and $I[f_i,f^{\rm UL},f^{\rm LC}]$ that 
leave enough room to accommodate smooth extrema based on a measure of the local 
curvature defined by 
\begin{equation}\label{eq:mp_M4}
 d^{\rm M4}_{i+\HALF} =
  \mm\left(4d_{i}   - d_{i+1}, 
           4d_{i+1} - d_{i}, d_i, d_{i+1}\right)\,,
\end{equation}
where $d_i = \Delta_{i+\HALF} - \Delta_{i-\HALF}$.
Using Eq. (\ref{eq:mp_M4}), one defines the median $f^{\rm MD}$ and the large curvature 
$f^{\rm LC}$ values as
\begin{equation}
 f^{\rm MD}_{i+\HALF} = \frac{f_{i} + f_{i+1}}{2} -
               \frac{1}{2}d^{\rm M4}_{i+\HALF}
\,,\quad
 f^{\rm LC}_{i+\HALF} = f_i + \frac{1}{2}\Delta_{i-\HALF}
                            + \frac{4}{3}d^{\rm M4}_{i-\HALF}\,,
\end{equation}
respectively.
The curvature measure provided by (\ref{eq:mp_M4}) is somewhat heuristic
and chosen to reduce the amount of room for local extrema to develop. 
The reconstruction illustrated preserves monotonicity and does not 
degenerate to first-order in proximity of smooth extrema.

%%%%%%%%%%%%%%%%%%%%%%%%%%%%%%%%%%%%%%%%%%%%%%%%%%%%%%%%%%%%%%%%%
\section{Numerical Tests}
\label{sec:numtest}
%
%
%
%%%%%%%%%%%%%%%%%%%%%%%%%%%%%%%%%%%%%%%%%%%%%%%%%%%%%%%%%%%%%%%%

In this section we present a series of test problems aimed at the 
verification of the FD methods previously described. 
The selected algorithms have been implemented in the PLUTO 
code for astrophysical gas-dynamics \cite{Mignone07} in order 
to ease inter-scheme comparisons through a flexible common computational 
framework. 

Unless otherwise stated, the specific heat ratio will be set to $\Gamma = 5/3$ and 
the Courant number $C_a$ will be taken equal to $0.8$, $0.4$ or $0.3$ 
for one, two and three dimensional computations, respectively.
%
%The time step, for test problems that include a resolution study, will
%be limited so as not to introduce the temporal error in that of the scheme.
%Since temporal integration is done using a Runge-Kutta $3$ method, the timestep
%will be taken equal to $\delta t = \delta t_{max} \times (n_0/n)^{Order/3}$, 
%where $\delta t_{max}$ is the maximum time step for the lowest resolution 
%used ($n_0$), while n is the current resolution. The ``Order'' is either $3$ or 
%$5$ for the LIMO3, WENO3 and MP5, WENO-Z respectively.
%
Errors for a generic flow quantity $Q$ are computed using the $L_1$ discrete 
norm defined by
\begin{equation}\label{eq:L1}
 \epsilon_1(Q) = \frac{1}{N_xN_yN_z}\sum_{i,j,k}\left|Q_{i,j,k} - Q_{i,j,k}^{\rm ref}\right| 
   \,,
\end{equation}
where the summation extends to all grid zones, $N_x, N_y$ and $N_z$ are the 
number of grid points in the three directions and $Q^{\rm ref}$ is a 
reference solution.
The divergence of magnetic field is quantified using Eq. (\ref{eq:L1}) with 
$\nabla\cdot\vec{B}$ computed as
\begin{equation}\label{eq:divBnum}
 \nabla\cdot\vec{B}= 
  \frac{B_{x,i+\HALF} - B_{x,i-\HALF}}{\Delta x} + 
  \frac{B_{y,j+\HALF} - B_{y,j-\HALF}}{\Delta y} +
  \frac{B_{z,k+\HALF} - B_{z,k-\HALF}}{\Delta z} \,,
\end{equation}
where the interface values are obtained through Eq. (\ref{eq:bx_intrfc}).

%%%%%%%%%%%%%%%%%%%%%%%%%%%%%%%%%%%%%%%%%%%%%%%%%%%%%%%%%%%%%%%%%
\subsection{Propagation of Circularly polarized Alfv\'en Waves}
\label{sec:alfv}
%
%
%%%%%%%%%%%%%%%%%%%%%%%%%%%%%%%%%%%%%%%%%%%%%%%%%%%%%%%%%%%%%%%%%

We start by considering a planar, circularly polarized Alfv\'en wave 
propagating along the $x$ direction. 
As the wave propagates, density and pressure stay constant whereas 
transverse vector components trace circles without changing their 
magnitude.
Denoting with $\omega$ and $k$ the angular frequency and wavenumber,
respectively, one has
\begin{equation}\label{eq:alfv_1D}
  \left(\begin{array}{c}
  v_x  \\ \noalign{\medskip}
  v_y  \\ \noalign{\medskip}
  v_z
\end{array}\right)
=
 \left( \begin{array}{c}
  v_{0x} \\ \noalign{\medskip}
  v_{0y} + A\sin\phi \\ \noalign{\medskip}
  v_{0z} + A\cos\phi 
\end{array}\right)
\,,\quad
  \left(\begin{array}{c}
  B_x  \\ \noalign{\medskip}
  B_y  \\ \noalign{\medskip}
  B_z  
\end{array}\right)
=
 \left( \begin{array}{c}
  c_a\sqrt{\rho} \\ \noalign{\medskip}
  \mp\sqrt{\rho}A\sin\phi \\ \noalign{\medskip}
  \mp\sqrt{\rho}A\cos\phi 
\end{array}\right)\,,
\end{equation}
where $\phi = kx - \omega t$, $\omega/k = v_{0x}\pm c_a$
is the corresponding phase velocity ($c_a$ is the Alfv\'en 
speed) and $A$ is the wave amplitude. 
The plus or minus sign corresponds to right or left propagating waves, 
respectively.
Here we consider a standing wave for which one has $v_{0x}=v_{0y}=v_{0z}=0$
and further set $\rho = 1$, $c_a=1$.

The one-dimensional solution given by (\ref{eq:alfv_1D}) is first
rotated by an angle $\gamma$ around the $y$ axis and subsequently 
by an angle $\alpha$ around the $z$ axis, as in \cite{MT10}.
The resulting transformation leaves scalar quantities 
invariant and produces vector rotations $\vec{q}\to\tens{R}_{\gamma\alpha}\vec{q}$, 
where $\vec{q}$ is either velocity or magnetic field and 
\begin{equation}\label{eq:rot_mat}
 \tens{R}_{\gamma\alpha} =
 \left(\begin{array}{ccc}
  \cos\alpha\cos\gamma &  -\sin\alpha & -\cos\alpha\sin\gamma  \\ \noalign{\medskip}
  \sin\alpha\cos\gamma &   \cos\alpha & -\sin\alpha\sin\gamma  \\ \noalign{\medskip}
  \sin\gamma           &   0          & \cos\gamma
 \end{array}\right)\,,\quad
 \tens{R}^{-1}_{\gamma\alpha} = 
 \left(\begin{array}{ccc}
  \cos\alpha\cos\gamma &   \sin\alpha\cos\gamma & \sin\gamma  \\ \noalign{\medskip}
  -\sin\alpha          &   \cos\alpha           &  0          \\ \noalign{\medskip}
 -\cos\alpha\sin\gamma & -\sin\alpha\sin\gamma  & \cos\gamma   
 \end{array}\right)\,,
\end{equation}
are the rotation matrix and its inverse.

Note that the rotation can be equivalently specified by prescribing 
the orientation of the wave vector $\vec{k}=(k_x,k_y,k_z)$ in a 
three-dimensional Cartesian frame  through the angles $\alpha$ and 
$\beta$ such that
\begin{equation}
  \tan\alpha = \frac{k_y}{k_x} \,,\quad
  \tan\beta  = \frac{k_z}{k_x} \,,\quad
\end{equation}
such that $\tan\gamma=\cos\alpha\tan\beta$.
With these choices, $\phi$ in (\ref{eq:alfv_1D}) becomes 
$\phi = \vec{k}\cdot\vec{x} - \omega t$ where 
$\omega = \pm|\vec{k}|$.

Periodicity is guaranteed by setting, without loss of generality, 
$k_x = 2\pi$ and by choosing the computational domain 
$x\in [0,1]$, $y\in[0,1/\tan\alpha]$ and $z\in[0,1/\tan\beta]$.
With these definitions the wave returns into the original position 
after one period 
\begin{equation}\label{eq:alfv_period}
 T = \frac{1}{\sqrt{1 + \tan^2\alpha + \tan^2\beta}}
\end{equation}
Different configurations can be specified in terms of 
the four parameters $\alpha,\beta, A$ and $p_0$ (background pressure).
One and two dimensional propagation are recovered 
by setting $\alpha=\beta=0$ and $\beta=0$, respectively.

%%%%%%%%%%%%%%%%%%%%%%%%%%%%%%%%%%%%%%%%%%%%%%%%
\subsubsection{One Dimensional Propagation}
\label{sec:alfv1D}
%
%
%%%%%%%%%%%%%%%%%%%%%%%%%%%%%%%%%%%%%%%%%%%%%%%%

\begin{table}[!ht]
\caption{Accuracy analysis for the one dimensional (third and fourth columns)
         and three dimensional (fifth and sixth) Alfv\'en wave propagation
         after one wave period.
         Errors are computed as
         $\sqrt{\epsilon_1(B_x)^2 + \epsilon_1(B_y)^2 + \epsilon_1(B_z)^2}$.
         The numerical scheme and the number of points $N_x$ in the $x$ direction
         are given in the first and second columns.
         For 3-D propagation, the resolution in the $y$ and $z$ direction 
         is set by $N_y=N_z=N_x/2$.}
\label{tab:alfv}
\centering
\begin{tabular*}{\textwidth}{@{\extracolsep{\fill}} lr rrrr}\hline
        &   &  \multicolumn{2}{c}{One Dimension}  &  \multicolumn{2}{c}{Three Dimensions} \\
  \cline{3-4} \cline{5-6}
Method & $N_x$  & $\epsilon_1\left(|\vec{B}|\right)$  & ${\cal O}_{L_1}$
                & $\epsilon_1\left(|\vec{B}|\right)$  & ${\cal O}_{L_1}$ \\ 
 \noalign{\smallskip} \hline\noalign{\smallskip}
\hline
WENO+3
 &  16 &       3.45E-03 &  -     &       2.54E-02 &  -      \\ \noalign{\smallskip} 
 &  32 &       4.39E-04 &   2.97 &       3.68E-03 &   2.79  \\ \noalign{\smallskip} 
 &  64 &       5.52E-05 &   2.99 &       4.47E-04 &   3.04  \\ \noalign{\smallskip} 
 & 128 &       6.91E-06 &   3.00 &       5.51E-05 &   3.02  \\ \noalign{\smallskip} 
 & 256 &       8.64E-07 &   3.00 &       6.85E-06 &   3.01  \\ \noalign{\smallskip} 
\hline \Lim
 &  16 &       3.36E-03 &  -     &       2.82E-02 &  -      \\ \noalign{\smallskip} 
 &  32 &       4.36E-04 &   2.95 &       3.76E-03 &   2.91  \\ \noalign{\smallskip} 
 &  64 &       5.53E-05 &   2.98 &       4.34E-04 &   3.11  \\ \noalign{\smallskip} 
 & 128 &       6.91E-06 &   3.00 &       5.46E-05 &   2.99  \\ \noalign{\smallskip} 
 & 256 &       8.65E-07 &   3.00 &       6.84E-06 &   3.00  \\ \noalign{\smallskip} 
\hline WENO-Z
 &  16 &       7.50E-04 &  -     &       4.10E-03 &  -      \\ \noalign{\smallskip} 
 &  32 &       2.40E-05 &   4.96 &       1.32E-04 &   4.96  \\ \noalign{\smallskip} 
 &  64 &       7.55E-07 &   4.99 &       3.89E-06 &   5.09  \\ \noalign{\smallskip} 
 & 128 &       2.36E-08 &   5.00 &       1.20E-07 &   5.02  \\ \noalign{\smallskip} 
 & 256 &       7.37E-10 &   5.00 &       3.74E-09 &   5.00  \\ \noalign{\smallskip} 
\hline MP5
 &  16 &       7.38E-04 &  -     &       3.41E-03 &  -      \\ \noalign{\smallskip} 
 &  32 &       2.40E-05 &   4.94 &       1.19E-04 &   4.84  \\ \noalign{\smallskip} 
 &  64 &       7.55E-07 &   4.99 &       3.81E-06 &   4.97  \\ \noalign{\smallskip} 
 & 128 &       2.36E-08 &   5.00 &       1.20E-07 &   4.99  \\ \noalign{\smallskip} 
 & 256 &       7.37E-10 &   5.00 &       3.74E-09 &   5.00  \\ \noalign{\smallskip} 
\hline
\end{tabular*}
\end{table}

As a first test, we consider one-dimensional propagating 
waves on the segment $x\in[0,1]$ using $N_x=2^q$ grid 
points with $q=4,\dots,8$.
We set the background pressure to be $p_0=0.1$ and the wave 
amplitude $A=0.1$.
The GLM correction is not necessary and has turned off for one 
dimensional propagation.

\begin{figure}[!h]
\centering
\includegraphics[width=0.485\textwidth]{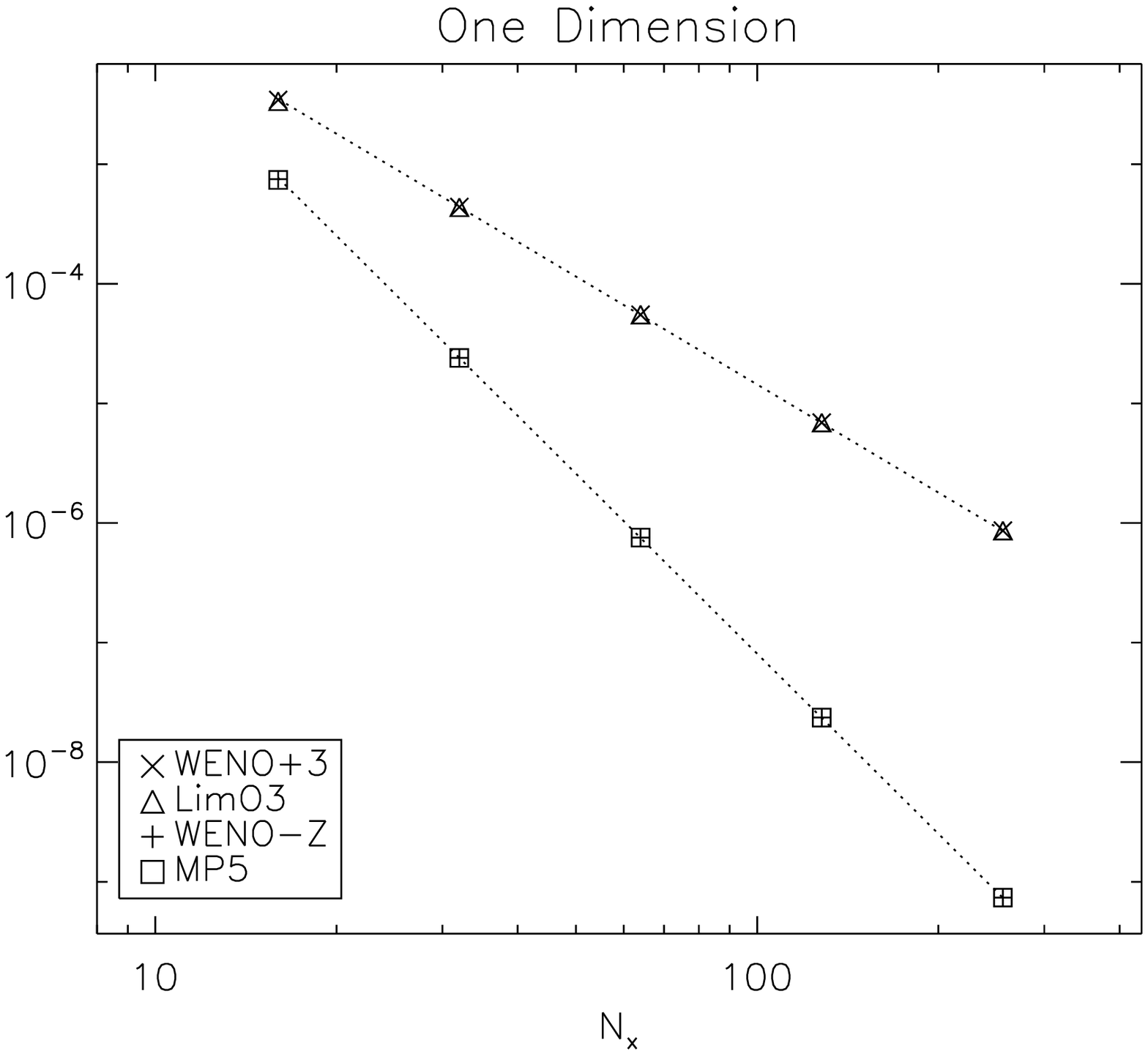}%
\includegraphics[width=0.485\textwidth]{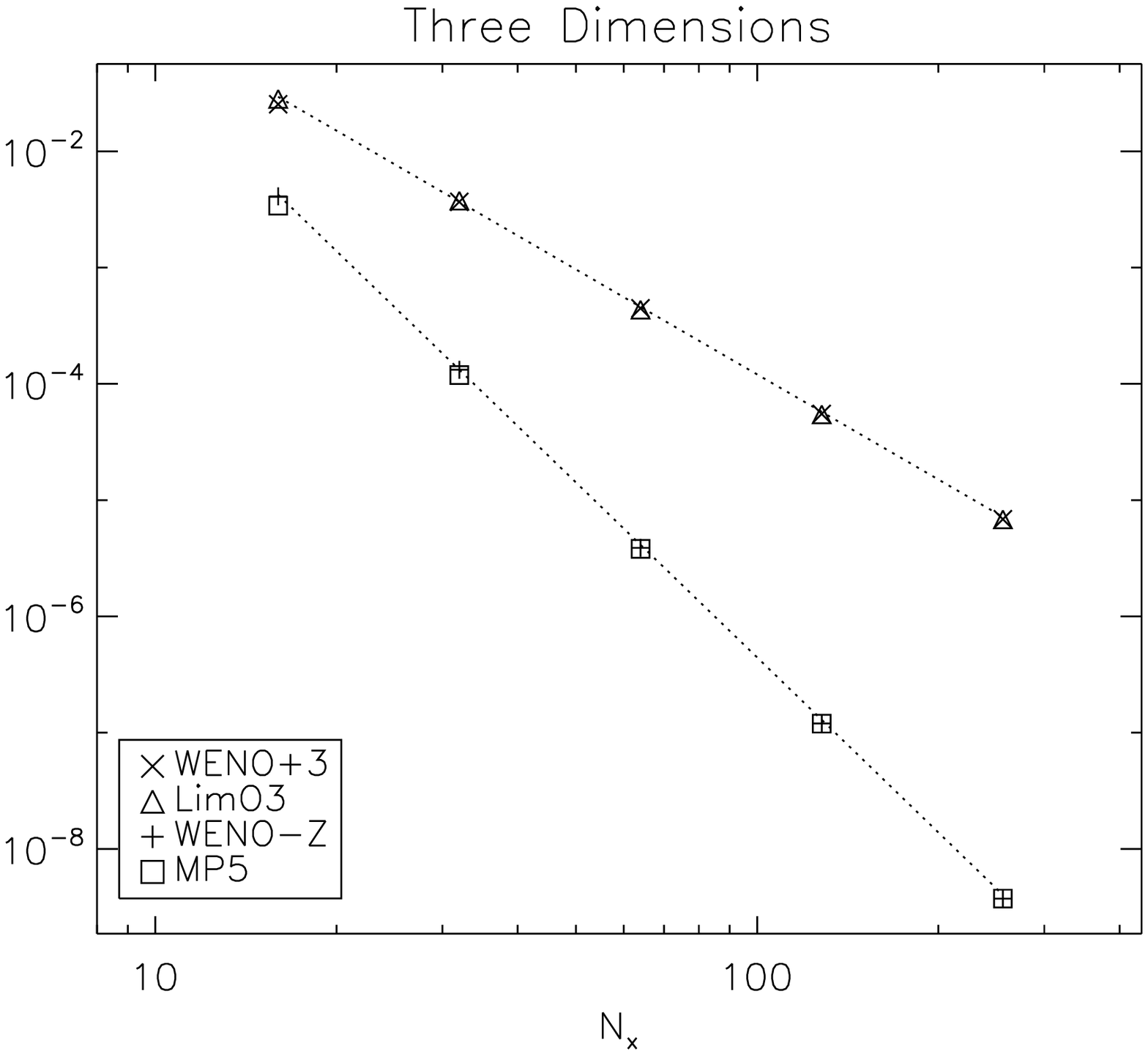}
 \caption{$L_1$ norm errors computed for the one-dimensional Alfv\'en wave 
          propagation (left panel) and the rotated three-dimensional 
          version (right panel).
          The cross, triangle, plus sign and square symbols refer to 
          computations carried out with WENO+3, \Lim, WENO-Z and MP5, 
          respectively, at the resolution $16, 32, 64, 128$ and $256$ 
          points using a CFL number of $0.8$ (in 1D) and $0.3$ (in 3D).
          The dotted lines gives the ideal convergence slope, that is, 
          $\propto \Delta x^3$ and $\propto \Delta x^5$, respectively. }
 \label{fig:alfv}
\end{figure}

In order to investigate the convergence of solution, the 
integration time step is adjusted to 
\begin{equation}\label{eq:dt_adj}
  \Delta t_{N} = \Delta t_{N_0}\left(\frac{N_0}{N}\right)^{r/3}
\end{equation}
where $\Delta t_{N_0}$ is the nominal time increment at the minimum 
resolution $N_0$, whereas $r\ge 3$ is the spatial accuracy of the scheme.
Errors (in $L_1$ norm) for the four selected schemes are plotted after one
wave period $T=1$ in the left panel of Fig \ref{fig:alfv} 
and arranged, together with the corresponding order of convergence,
in the third and fourth columns of Table \ref{tab:alfv}.
All schemes meet the expected order of accuracy (i.e. 3 for
\Lim and WENO+3, 5 for WENO-Z and MP5) with no significant differences.
It is remarkable that, at the resolution of $64$ zones,
the fifth-order schemes achieve essentially the same accuracy as the 
third-order schemes that make use of four times (i.e. $N_x=256$) 
as many points.

%%%%%%%%%%%%%%%%%%%%%%%%%%%%%%%%%%%%%%%%%%%%%%%%%%%%%%%%%
\subsubsection{Three Dimensional Oblique Propagation}
\label{sec:alfv3D}
%
%
%%%%%%%%%%%%%%%%%%%%%%%%%%%%%%%%%%%%%%%%%%%%%%%%%%%%%%%%%

A three dimensional configuration is obtained by rotating the
one-dimensional setup described in \S\ref{sec:alfv1D} by the angles
$\alpha=\beta=\tan^{-1}2$ so that $\tan\gamma=2/\sqrt{5}$ in 
Eq. (\ref{eq:rot_mat}).
The background pressure is $p_0=0.1$ and the wave has amplitude 
$A=0.1$.
The size of the computational box turns out to be $x\in[0,1]$, $y\in[0,1/2]$,
$z\in[0,1/2]$ and the number of grid points is set by $N_y = N_z = N_x/2$,
where $N_x$ changes as in \S\ref{sec:alfv1D}.
Integration lasts for one wave period, i.e., $t=T=1/3$ and 
the time step is determined by the same condition given by Eq. (\ref{eq:dt_adj}).
Thus, apart from the different normalization, our setup is identical to 
that used in \cite{GS08}.

Errors are plotted at different resolutions in the right panel of Fig
\ref{fig:alfv} and sorted in Table \ref{tab:alfv} for all schemes.
On average, errors are $\sim 4$ larger than their one-dimensional 
counterparts but the overall behavior meets the expected order of accuracy with 
MP5 and \Lim performing slightly better than WENO-Z and WENO+3, 
respectively.
As for the 1D case, roughly $1/4$ of the resolution is required by 
a fifth-order scheme to match the accuracy of a third-order one.

\begin{figure}[!h]
\centering
\includegraphics[width=\textwidth]{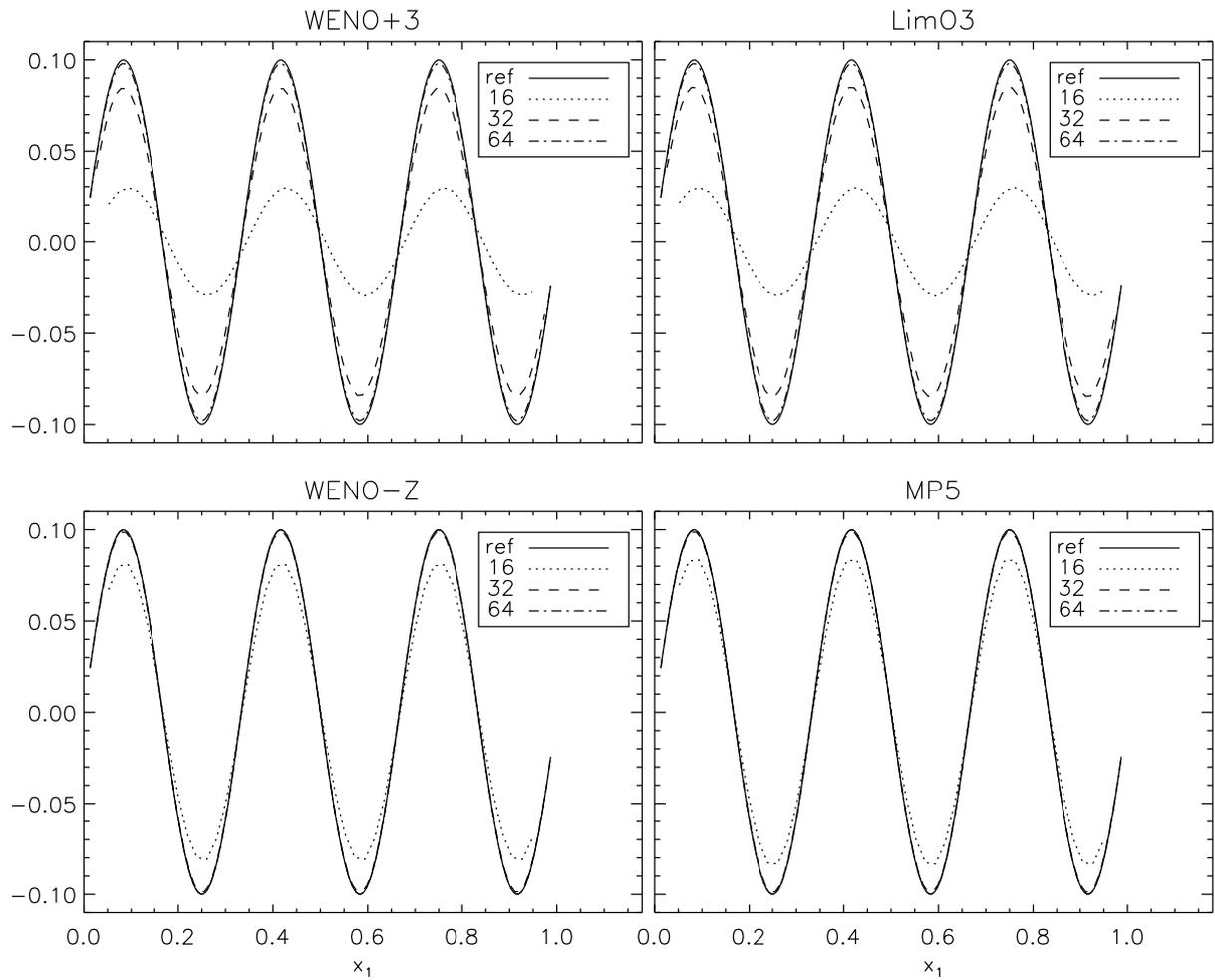}
 \caption{Scatter plots of the $y$ component of magnetic field in the 
          original one-dimensional frame at $t=5/3$, after 5 revolutions.
          Each panel plots every point of the three-dimensional array 
          $-B_x\sin\alpha + B_y\cos\alpha$ as a function of the longitudinal 
          coordinate $\vec{k}\cdot\vec{x}/|\vec{k}|$ along the direction 
          of wave propagation.
          The lack of scatter demonstrates that the algorithm retains 
          the expected planar symmetry.
          The solid line gives the reference solution at $t=0$ while
          dotted, dashed and dot-dashed lines corresponds to computations 
          carried with $N_x=16,32,64$ points, respectively.
          The CFL number was set to $C_a=0.3$.}
 \label{fig:scatter}
\end{figure}

Following \cite{GS08}, we construct in Fig \ref{fig:scatter} 
a scatter plot of the magnetic field component parallel to the $y$ 
axis of the original one dimensional frame.
This is achieved by plotting, for every point in the computational domain,
the $y$ component of $\tens{R}^{-1}_{\gamma\alpha}\vec{B}$ as a 
function of the normal ($x$) coordinate of 
$\tens{R}^{-1}_{\gamma\alpha}\vec{x}$, where
$\tens{R}_{\gamma\alpha}$ is the rotation matrix introduced
in (\ref{eq:rot_mat}).
The ability of the scheme to retain the planar symmetry during 
the computation is confirmed by the lack of scatter in the plots.
The profiles at different resolutions verify the general 
trend established in Table \ref{tab:alfv} and deviations from 
the exact solution appear to be imperceptible for $N_x > 64$ for the 
third-order schemes and already at $N_x\gtrsim 32$ for the
fifth-order schemes.
%This clearly demonstrates the superiority of the latter over the former.

Overally, the results obtained with third- and fifth-order accurate 
schemes outperform traditional TVD schemes, such as the 
CT-PPM algorithm of \cite{GS08} yielding at most second-order accurate 
solutions.
The CPU costs associated with WENO+3, \Lim, WENO-Z and MP5 show, for this 
test problem, a relative scaling $1:0.98:1.46:1.31$, respectively.

%%%%%%%%%%%%%%%%%%%%%%%%%%%%%%%%%%%%%%%%%%%%%%%%%%%%%%%%%%%%%%%%%%%%%%%%%%%%%%%%%%
\subsubsection{Numerical Dissipation and Long Term Decay in Two Dimensions}
\label{sec:alfv2D}
%
%
%%%%%%%%%%%%%%%%%%%%%%%%%%%%%%%%%%%%%%%%%%%%%%%%%%%%%%%%%%%%%%%%%%%%%%%%%%%%%%%%%%

As already stated, circularly polarized Alfv\'en waves are an 
exact nonlinear solution of the MHD equations and measuring their decay 
provides a direct indication of the intrinsic numerical viscosity and 
resistivity possessed by the underlying algorithm, see \cite{RJF95, Balsara04,BRDM09}. 
This study is relevant, for example, in the field of MHD turbulence 
modeling where one should carefully control the amount of 
directionally-biased dissipation introduced by waves propagating inclined 
to the mesh. 
The error introduced during an oblique propagation is usually minimized at 
$45^\circ$ since contributions coming from different directions have 
comparable magnitude.
On the contrary, waves propagating at smaller inclination angles make 
the problem more challenging.

Our setup builds on \cite{Balsara04} although we adopt a slightly 
different, more severe, configuration.
Using the notations introduced in \S\ref{sec:alfv}, we set 
$\tan\alpha = 6$, $\tan\beta=0$, $A=0.2$ and prescribe the background 
pressure to be $p_0 = 1$. The corresponding ratio of the plasma pressure 
to the (unperturbed) magnetic pressure is then given by 
$p/(2\rho c^2_a)=1/2$, where $c_a=1$ is the wave propagation
speed.
The choice of the inclination angle determines the computational domain 
$x\in[0,1]$, $y\in[0,1/6]$ as well as the wave period $T=1/\sqrt{37}$
from Eq. (\ref{eq:alfv_period}).
The final integration time $t=16.5$ is chosen by having the wave cross the 
domain $\approx 100$ times.
This configuration results in a more arduous test than \cite{Balsara04} 
where the wave period was $6\sqrt{4\pi}$ longer and the integration 
was stopped after $\approx 37$ wave transits.

\begin{figure}[!h]
\centering
\includegraphics[width=\textwidth]{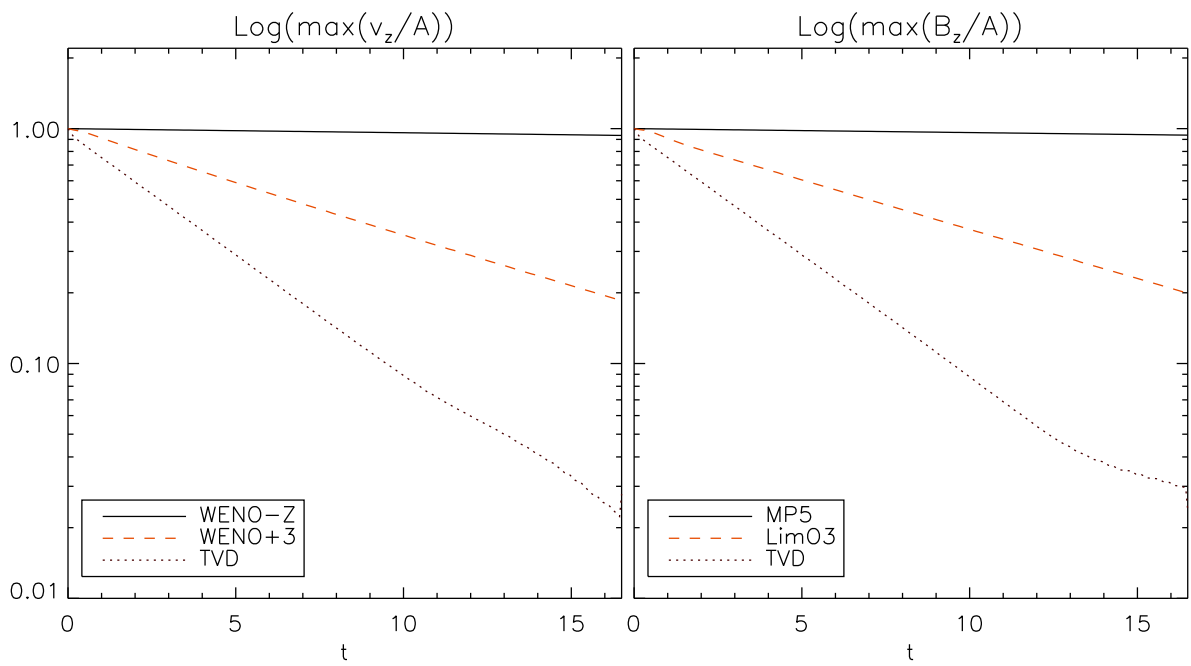}
 \caption{Long term decay of circularly polarized Alfv\'en waves after
         $16.5$ time units, corresponding to $\sim 100$ wave periods.
         In the left panel, we plot the maximum value of the vertical 
         component of velocity as a function of time for the WENO-Z 
         (solid line) and WENO+3 (dashed line) schemes. 
         For comparison, the dotted line gives the result obtained by 
         a second-order TVD scheme.
         The panel on the right shows the analogous behavior of the 
         vertical component of magnetic field $B_z$ for \Lim and MP5.
         For all cases, the resolution was set to $120\times 20$ and the 
         Courant number is $0.4$.}
 \label{fig:decay}
\end{figure}

Fig \ref{fig:decay} shows, at the resolution of $120\times 20$ mesh 
points, the maximum values of the vertical $z$ components of velocity 
(left panel) and magnetic field (right panel) as functions of time. 
By the end of the simulation, third-order schemes 
(dashed lines) show some degree of dissipation with the wave 
amplitude being reduced to $\sim 20$ per cent of its initial value.
On the contrary, schemes of order five (solid lines in the
figure) preserve the original shape more accurately and the amplitude 
retains $\sim 94$ per cent of its nominal value.
These results are compared, for illustrative purposes, to a
$2^{\rm nd}$ order TVD scheme using the Monotonized Central 
difference limiter (dotted lines), showing that the initial peak 
values have scaled down to $\sim 3$ per cent, thus 
showing a considerably larger level of numerical dissipation.

These results are in agreement with previous investigations 
\cite{Balsara04, BRDM09} and strongly supports the idea that problems involving 
complex wave interactions may benefit from using higher-order schemes such 
as the ones presented here.

%%%%%%%%%%%%%%%%%%%%%%%%%%%%%%%%%%%%%%%%%%%
\subsection{Shock tube problems}
%
%
%%%%%%%%%%%%%%%%%%%%%%%%%%%%%%%%%%%%%%%%%%%

Shock tube problems are commonly used to test the ability of the scheme in 
describing both continuous and discontinuous flow features. 
In the following we consider two and three dimensional 
rotated configurations of standard one dimensional tubes.
The default value for the parameter $\alpha_p$ controlling monopole damping 
(see Eq. \ref{eq:source_exp}) is $0.8$.

%%%%%%%%%%%%%%%%%%%%%%%%%%%%%%%%%%%%%%%%%%%
\subsubsection{Two-Dimensional Shock tube}
\label{sec:BW}
%
%
%%%%%%%%%%%%%%%%%%%%%%%%%%%%%%%%%%%%%%%%%%%

Following \cite{JW99, Li08}, we consider a rotated version of the Brio-Wu 
test problem \cite{BW88} with left and right states are given by 
\begin{equation}
\left\{
\begin{array}{lclr}
\vec{V}_L & = & \DS
 \left(1, 0, 0, 0.75, 1, 1\right)^T 
              & \mathrm{for}\quad  x_1 < 0 \,, \\ \noalign{\medskip}
\vec{V}_R & = & \DS \left(0.125, 0, 0, 0.75, -1, 0.1\right)^T 
              & \mathrm{for}\quad  x_1 > 0  \,,
\end{array}\right.
\end{equation}
where $\vec{V} = \left(\rho,v_1,v_2, B_1, B_2, p\right)$ is
the vector of primitive variables. The subscript
``1'' gives the direction perpendicular to the initial surface of discontinuity
whereas ``2'' corresponds to the transverse direction. 
Here $\Gamma = 2$ is used and the evolution is interrupted at time $t = 0.2$, 
before the fast waves reach the borders. 

In order to address the ability to preserve the initial planar symmetry we 
rotate the initial condition by the angle $\alpha=\pi/4$ in a two dimensional 
plane with $x\in[-1,1]$ and $y\in[-0.01,0.01]$ using  
$N_x\times N_x/100$ grid points, with $N_x=600$.
%We adopt the same configuration as in \cite{JW99} and choose the rotation 
%angle to be $\alpha = \pi/4$ making the discontinuity propagating 
%along the grid main diagonal. 
Vectors follow the same transformation given by Eq. (\ref{eq:rot_mat}) with 
$\beta=\gamma=0$.
This is known to minimize errors of the longitudinal component of the magnetic 
field (see for example the discussions in \cite{Toth00, GS05}).
Boundary conditions respect the translational invariance specified by the 
rotation: for each flow quantity we prescribe 
$q(i,j) = q(i\pm\delta i, j\pm\delta j)$ where $(\delta i,\delta j) = (1,-1)$, 
with the plus (minus) sign for the leftmost and upper (rightmost and lower) 
boundary. 
Computations are stopped before the fast rarefaction waves reach the boundaries, 
at $t=0.2\cos\alpha$.

\begin{figure}[!h]
\centering
\includegraphics[width=\textwidth]{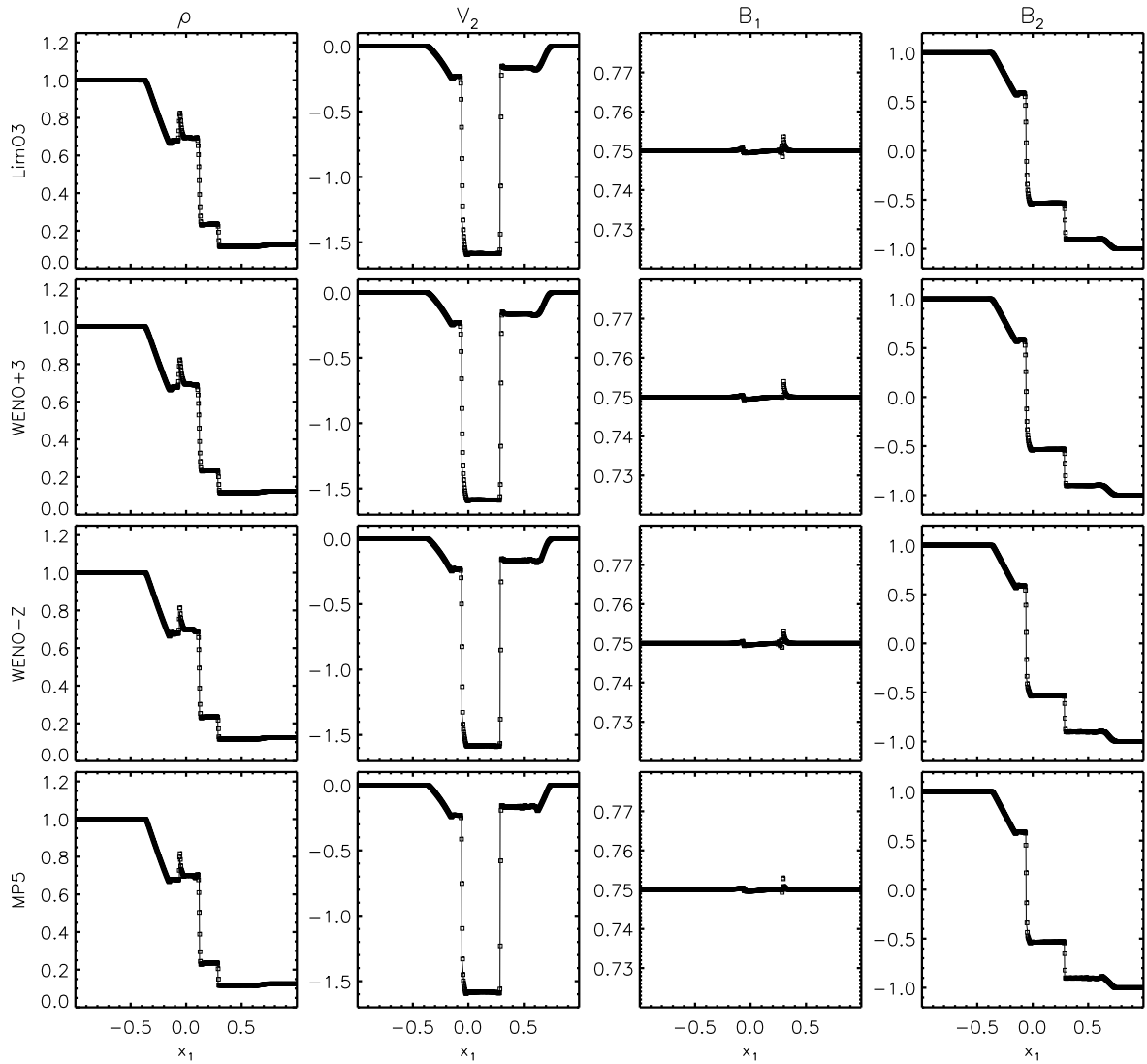}
 \caption{Primitive variable profiles for the 2D shock tube problem at 
  $t=0.2\cos\alpha=0.2/\sqrt{2}$, along the rotated direction $\rm x_1$. 
  From left to right: density, transverse velocity, 
  longitudinal and transverse magnetic field components are displayed.
  The mesh resolution is $600\times 6$ and the Courant number is $0.4$.
  Symbols correspond to the 2D computations whereas the solid lines 
  gives the reference solution. }
 \label{fig:sod2d}
\end{figure}

\begin{figure}\centering
\includegraphics[width=\textwidth]{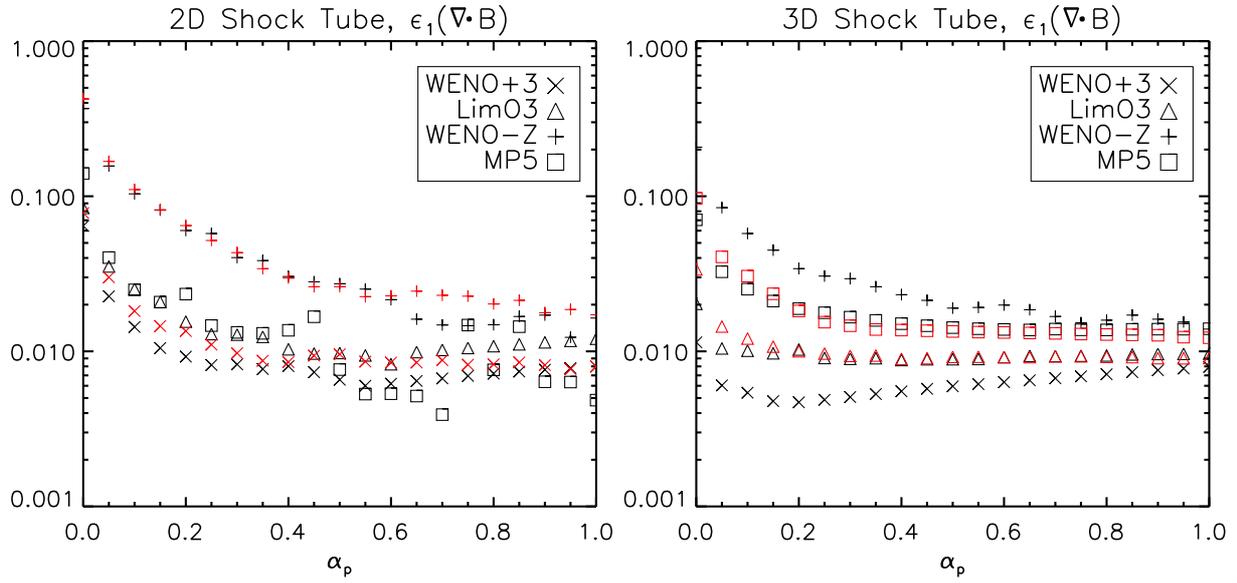}
 \caption{Divergence errors as function of the damping parameter $\alpha_p$
          for the shock tube problems in 2D (left, $t=0.2/\sqrt{2}$) and 
          3D (right, $t=0.8/\sqrt{21}$).
          Symbols in black color are used to distinguish between different 
          schemes at the nominal resolutions ($600\times 6$ in 2D and 
          $768\times 8\times 8$ in 3D), see the legend.
          Computations carried at twice the resolution ($1200\times 12$ in 
          2D and $1536\times 16\times 16$ in 3D) are shown using 
          symbols in red color.}
 \label{fig:sod_alpha}
\end{figure}

Fig \ref{fig:sod2d} shows the primitive variable profiles for all schemes 
against a one-dimensional reference solution obtained on a base grid of 
$1024$ zones with $5$ levels of refinement.
Errors in $L_1$ norm, computed with respect to the same reference solution, 
are sorted in Table \ref{table:Sod} for density and the normal 
component of magnetic field.
The out-coming wave pattern is comprised, from left to right, of a fast 
rarefaction, a compound wave (an intermediate shock followed by a slow 
rarefaction), a contact discontinuity, a slow shock and a fast rarefaction wave. 
We see that all discontinuities are captured correctly and the overall behavior 
matches the reference solution very well. 
The normal component of magnetic field is best described with MP5 and 
does not show erroneous jumps. 
Indeed, the profiles are essentially constant with small
amplitude oscillations showing a relative peak $\sim\,0.7\%$.
Divergence errors, typically $\lesssim 10^{-2}$, remain bounded with resolution 
and tend to saturate when the damping parameter $\alpha_p \gtrsim 0.4$ for
both 2 and 3D calculations, see Fig \ref{fig:sod_alpha}.
In this sense, our results favourably 
compare to those of \cite{JW99, Li08} and \cite{DBFM08}.

Fifth-order methods exhibit less dissipation across jumps, with fewer points in 
each discontinuous layer. 
Still, the accuracy gained from third to fifth-order accurate schemes 
(see Table \ref{table:Sod}) is only a factor $1.5-2$ since interpolation 
across discontinuities usually degenerates to lower-order to suppress spurious 
oscillations.

\begin{table}\centering
\caption{One dimensional $L_1$ norm errors for density, normal component of magnetic
         field and $|\nabla\cdot\vec{B}|$ for the two and three-dimensional shock tube.}
\label{table:Sod}
\centering

\begin{tabular*}{\textwidth}{@{\extracolsep{\fill}} lr rrrrr}\hline
        &   &  \multicolumn{2}{c}{Two Dimensions}  &  \multicolumn{2}{c}{Three Dimensions} &  \\
  \cline{2-4} \cline{5-7}
Method & $\epsilon_1(\rho)$  & $\epsilon_1(B_1)$ & $\epsilon_1\left(\nabla\cdot\vec{B}\right)$ & 
         $\epsilon_1(\rho)$  & $\epsilon_1(B_1)$ & $\epsilon_1\left(\nabla\cdot\vec{B}\right)$ \\ 
 \noalign{\smallskip} \hline\noalign{\smallskip}\hline
WENO+3
&      4.11E-03 &     8.53E-05 & 7.19E-03 &   1.82E-03  & 4.41E-05 & 7.12E-03  \\ \noalign{\smallskip} 
\Lim
&      3.61E-03 &     8.74E-05 & 1.08E-02 &   1.63E-03  & 4.07E-05 & 9.41E-03  \\ \noalign{\smallskip} 
WENO-Z
&      2.72E-03 &     7.90E-05 & 1.48E-02 &   1.29E-03  & 5.41E-05 & 1.59E-02  \\ \noalign{\smallskip} 
MP5
&      2.31E-03 &     6.24E-05 & 7.60E-03 &   1.07E-03  & 2.22E-05 & 1.37E-02  \\ \noalign{\smallskip} 
\hline
\end{tabular*}
\end{table}

%%%%%%%%%%%%%%%%%%%%%%%%%%%%%%%%%%%%%%%%%%%%%%%%
\subsubsection{Three-Dimensional Shock tube}
%
%
%%%%%%%%%%%%%%%%%%%%%%%%%%%%%%%%%%%%%%%%%%%%%%%%

The second Riemann problem was introduced by \cite{RJ95} and later
considered by \cite{RJF95,Toth00,BS00} and by \cite{GS08, MT10} in 
3D.
The primitive variables are initialized as 
\begin{equation}\label{eq:ic_3dst}
\left\{\begin{array}{lclr}
\vec{V}_L & = & \DS
 \left(1.08,1.2,0.01,0.5,\frac{2}{\sqrt{4\pi}},\frac{3.6}{\sqrt{4\pi}} 
                         \frac{2}{\sqrt{4\pi}},0.95\right)^T 
              & \mathrm{for}\quad  x_1 < 0  \\ \noalign{\medskip}
\vec{V}_R & = & \DS \left(1,0,0,0,\frac{2}{\sqrt{4\pi}},\frac{4}{\sqrt{4\pi}},
                                  \frac{2}{\sqrt{4\pi}},1\right)^T 
              & \mathrm{for}\quad  x_1 > 0  
\end{array}\right.
\end{equation}
where $\vec{V} = \left(\rho,v_1,v_2, v_3, B_1, B_2, B_3, p\right)$.
A reference solution at $t=0.2$ is obtained on the domain $x\in[-0.75,0.75]$ 
using $2048$ grid points and $5$ levels of refinement.
Our setup draws on the three dimensional version of \cite{GS08} and \cite{MT10} 
where the initial condition (\ref{eq:ic_3dst}) is rotated using 
Eq. (\ref{eq:rot_mat}) by the angles $\alpha$ and $\gamma$ such that 
$\tan\alpha=-1/2$ and $\tan\gamma=1/(2\sqrt{5})$ (corresponding to 
$\tan\beta = 1/4$). 
With this choice the planar symmetry is respected by an integer shift of cells.
The computational domain consists of $768\times 8\times 8$ zones and spans 
$[-0.75,0.75]$ in the $x$ direction while $y,z\,\in [0,0.015625]$.
Computations stop at $t=0.2\cos\alpha\cos\gamma$ (note the misprint in 
\cite{MT10}).

\begin{figure}[!h]\centering
\includegraphics[width=\textwidth]{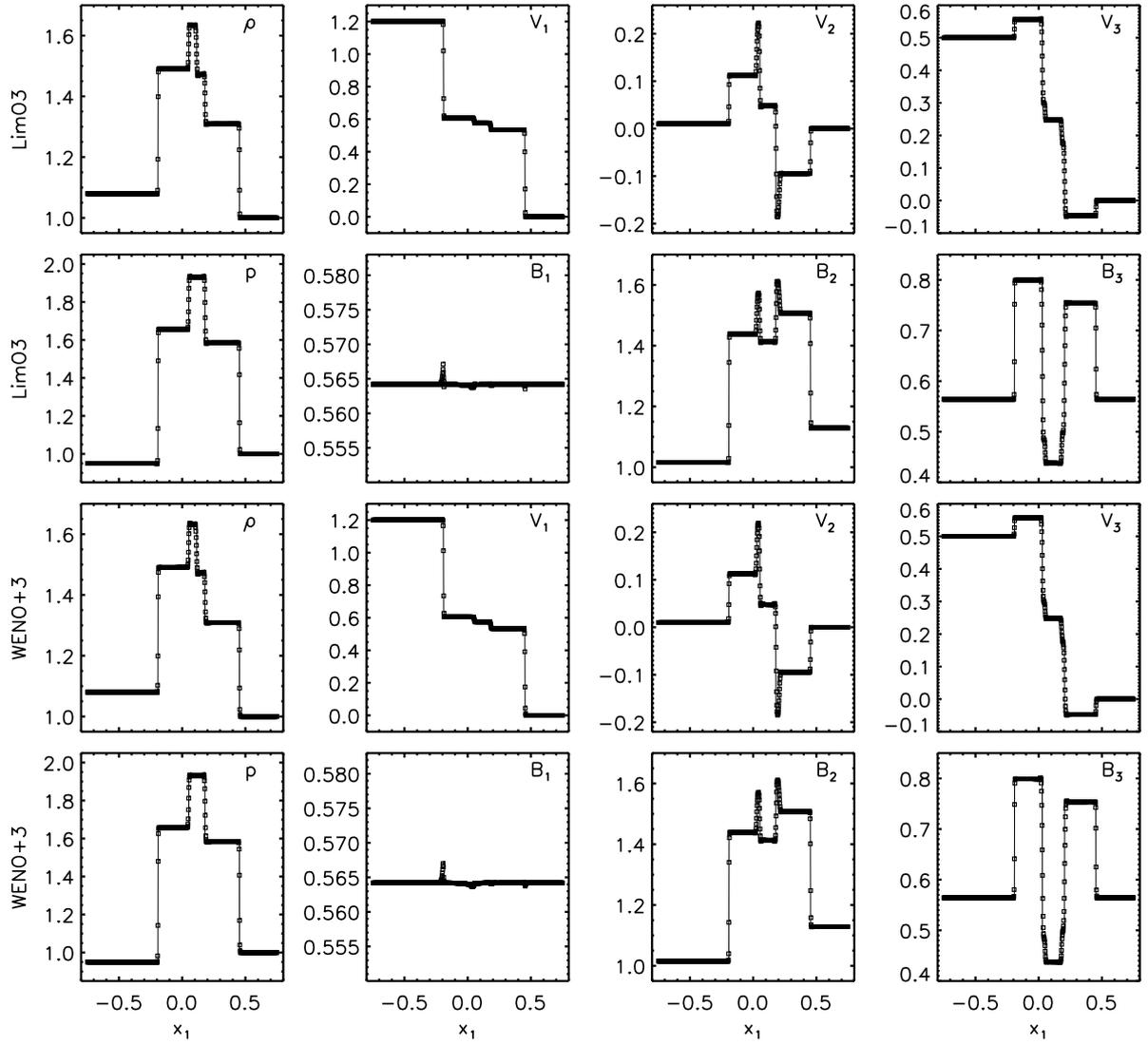}
 \caption{Primitive variable profiles for the 3D shock tube problem at 
  $t=0.2\cos\alpha\cos\gamma=0.8/\sqrt{21}$ obtained with the third-order schemes.
  Density, pressure, velocity and magnetic field components parallel
  and transverse to the direction of propagation are plotted as functions 
  of the longitudinal component $x_1$. 
  The mesh resolution is $768\times 8\times 8$ and the Courant number
  is $0.3$.}
 \label{fig:sod3d_a}
\end{figure}

\begin{figure}\centering
\includegraphics[width=\textwidth]{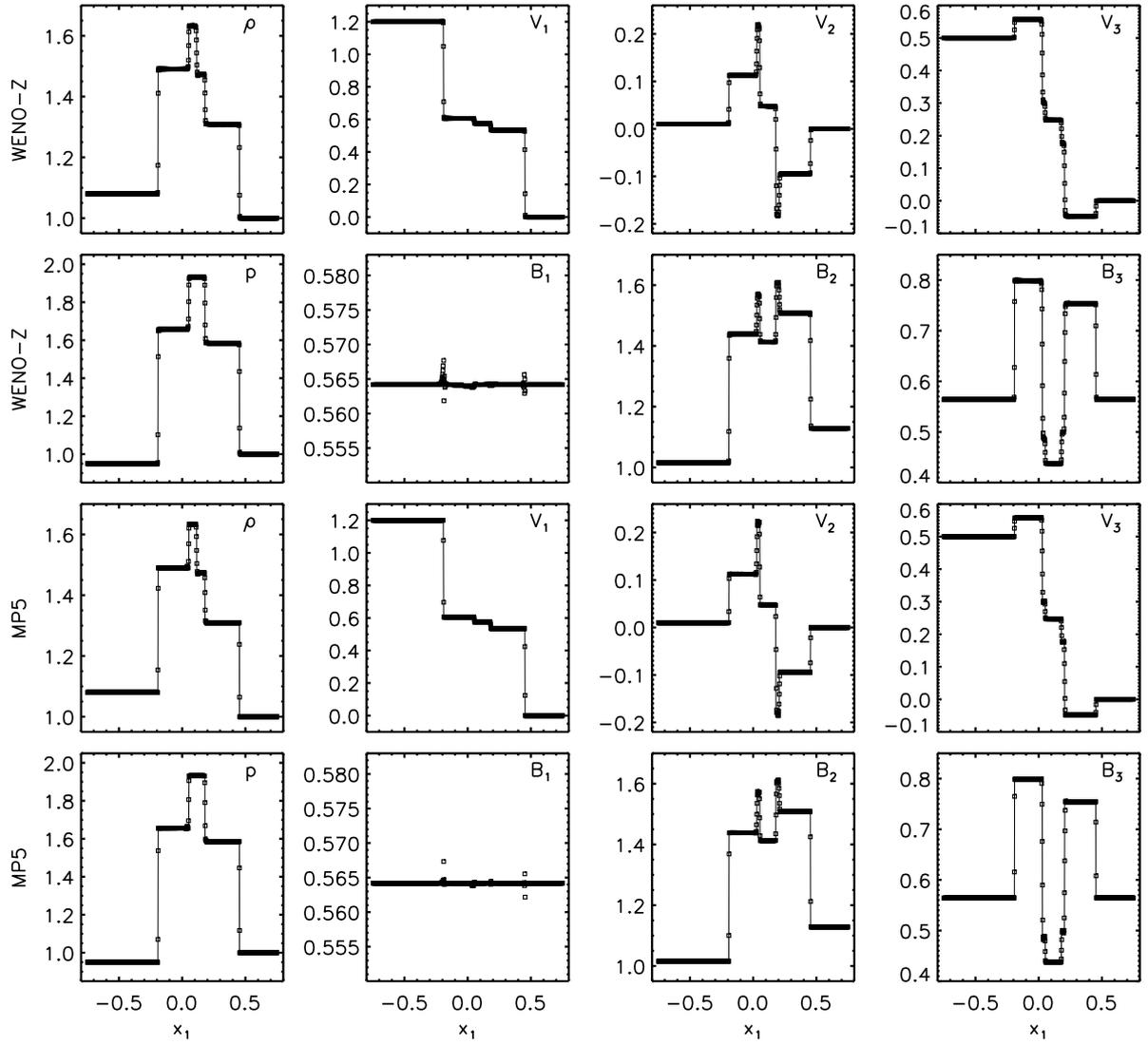}
 \caption{Same as Fig \ref{fig:sod3d_a} but for the fifth-order schemes
          WENO-Z and MP5.}
 \label{fig:sod3d_b}
\end{figure}

Fig \ref{fig:sod3d_a} and \ref{fig:sod3d_b} show primitive variable profiles 
obtained with third- and fifth-order schemes, respectively. 
The wave pattern consists of a contact discontinuity that separates 
two fast shocks, two slow shocks and a pair of rotational discontinuities. 
Table \ref{table:Sod} confirms again that the gain from high-order
methods is not particularly significant when the flow is discontinuous. 
Our results favorably compare with those of other investigators
and no prominent over/under-shoots are observed. 
Moreover, the amount of oscillations in the normal component of 
the magnetic field is comparable to (or smaller than) those found 
in \cite{GS08,MT10} and
divergence errors behave in a very similar way to the 2D case 
(see also the right panel in Fig \ref{fig:sod_alpha}).

The computational costs relative to that of WENO+3 ($=1$) are
found, for this problem, to be $0.99:1.48:1.25$ for \Lim, WENO-Z and MP5, respectively. 

%%%%%%%%%%%%%%%%%%%%%%%%%%%%%%%%%%%%%%%%%%%%%%%%%%%%%%%%%%%%%%%%%%%%%%%%%%%%%%%%
\subsection{Iso-density MHD Vortex advection}
%
%
%    Paper               2D                 Error         3D        Error
%
%  Balsara04            YES                momentum,     -           -
%                       (TVD)              B vectors  
%
%  Balsara09           YES,                 Bx           -          -
%                      (High order FV)
%
%  Balsara et al 09     YES                Bx             -         -
%                      (High order FV)
%  Dumbser et al 09      -                               YES         Bx
%                                                       (DG, FV)
%%%%%%%%%%%%%%%%%%%%%%%%%%%%%%%%%%%%%%%%%%%%%%%%%%%%%%%%%%%%%%%%%%%%%%%%%%%%%%%%

The following problem has been introduced in \cite{Balsara04} and 
lately considered by \cite{BRDM09, Balsara09, DBFM08}. 
The initial condition, satisfying the time-independent MHD equations,
consists of a magnetized vortex structure in force equilibrium 
that propagates along the main diagonal of the computational box (a square 
in 2D and a cube in 3D). Here we set $\alpha_p=0.4$.

%%%%%%%%%%%%%%%%%%%%%%%%%%%%%%%%%%%%%%%%%%%%%%%%%%%%%%%%%%%%%%%%
\subsubsection{Two Dimensional Propagation}
%
%%%%%%%%%%%%%%%%%%%%%%%%%%%%%%%%%%%%%%%%%%%%%%%%%%%%%%%%%%%%%%%%

Following Dumbser et al. \cite{DBFM08}, we perform computations on 
the Cartesian box $[-5,5]^2$ with an initial flow described by
$\rho = 1$, $\vec{v} = \vec{1} +(- y,x,0)\,\kappa {\rm e}^{q\,(1-r^2)}$, 
$\vec{B} = (- y,x,0)\,\mu {\rm e}^{q\,(1-r^2)}$ and 
$p = 1 + 1/(4\,q)\,(\mu^2\,(1 - 2\,q\,r^2) - \kappa^2\,\rho)\,{\rm e}^{2\,q\,(1-r^2)}$.
The constants $\kappa$ and $\mu$ are chosen to be equal to $1/2\pi$ 
while $r = \sqrt{x^2+y^2}$. The simulations are evolved
for 10 time units with periodic boundary conditions, i.e. a single passage 
of the vortex through the domain. The parameter $q$ is chosen equal to $0.5$ for 
third-order schemes, effectively reproducing the configuration shown in 
\cite{Balsara04, Balsara09}.
For WENO-Z and MP5, on the other hand, we choose $q=1$ in order to reduce the 
unwanted effects produced by the small jump in the magnetic field at the periodic 
boundaries, as argued in \cite{DBFM08}.

\begin{figure}\centering
\includegraphics[width=0.45\textwidth]{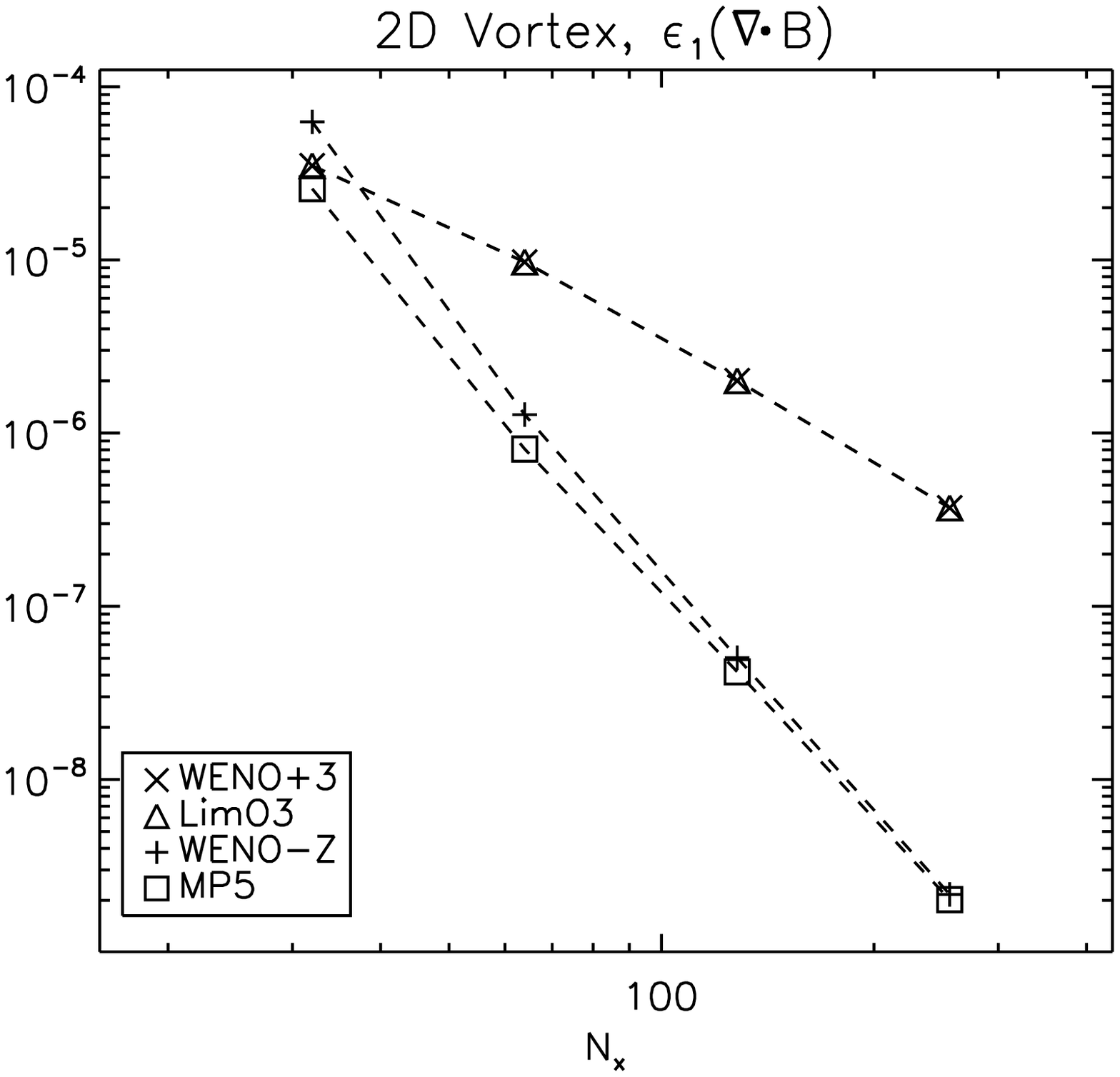}
\includegraphics[width=0.45\textwidth]{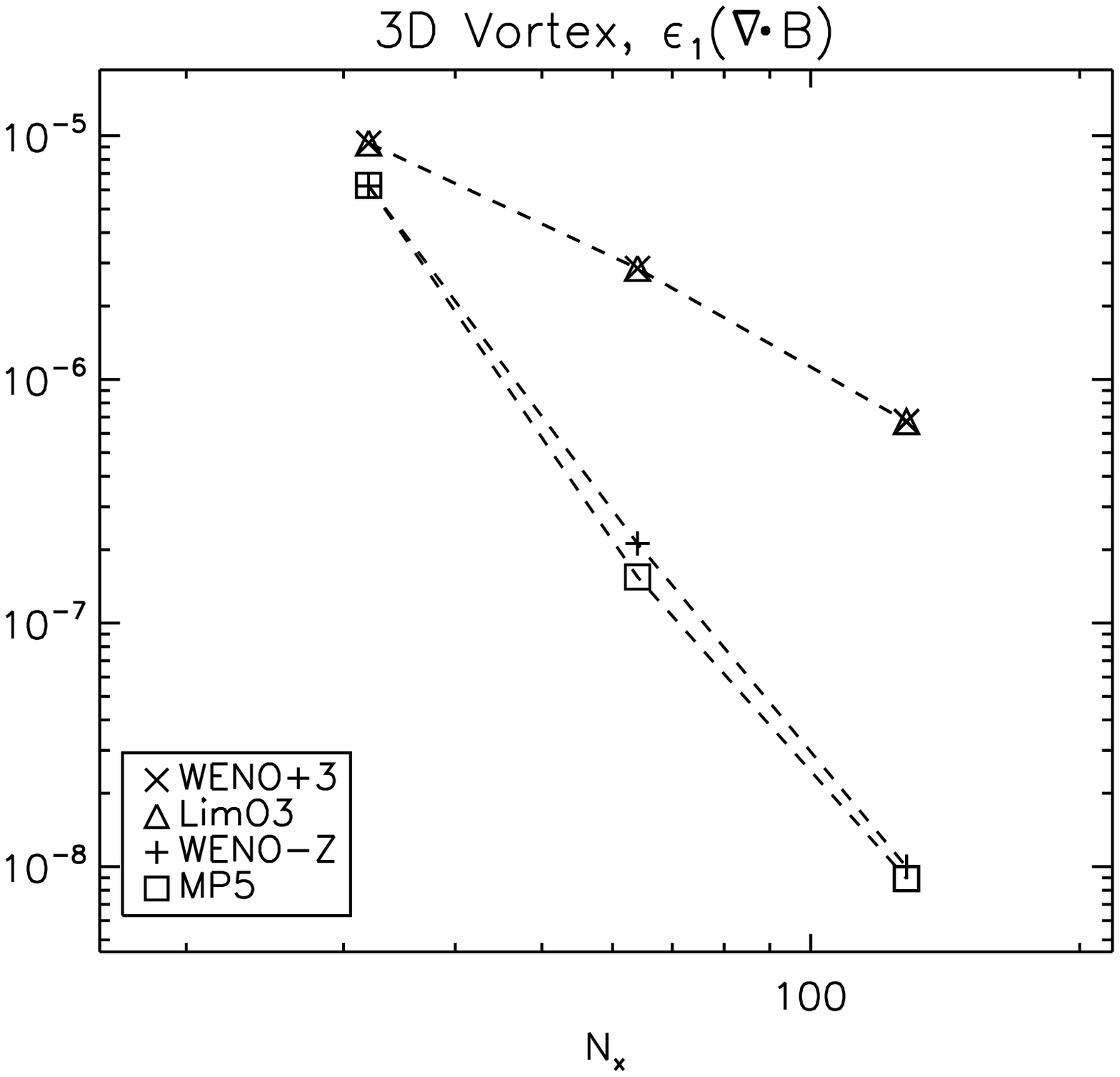}
 \caption{$L_1$ norm error of the divergence of magnetic field as functions 
  of the resolution ($N_x$) for the 2D (left panel) and 3D (right panel) 
  vortex problems at $t=10$.
  Different symbols corresponds to the selected reconstruction algorithms.}
 \label{fig:vortex}
\end{figure}

In order to compare our results to the findings of the latter study, 
we report, in Table \ref{table:L1_vortex}, errors for $B_x$ measured both in 
$L_1$ and $L_2$ norms and the corresponding convergence rates.
All schemes quickly converge to the asymptotic order of accuracy.
Remarkably, errors obtained with the third-order schemes are identical 
and somewhat better than those of \cite{Balsara09}.
At the resolution of $128^2$, fifth-order schemes yield errors 
$\sim 4$ times smaller than third-order ones at $256^2$.
A comparison between third- and fifth-order schemes from Fig \ref{fig:vortex}
reveals that divergence errors rapidly decrease with resolution 
following a similar pattern.
This eloquently advocates towards the use of higher-order schemes.

\begin{table}\centering
\caption{$L_1$ and $L_2$ norm errors and corresponding convergence rates 
         for the MHD Vortex problem in 2D (columns 3-6) and 3D (columns 7-10)
         at $t=10$.}
\label{table:L1_vortex}
\centering
\begin{tabular*}{\textwidth}{@{\extracolsep{\fill}} cc cccccccccc}\hline
        &   &  \multicolumn{4}{c}{Two Dimensions}  &  \multicolumn{4}{c}{Three Dimensions}   \\
  \cline{3-6} \cline{7-10}
Method & $N_x$  & $\epsilon_1(B_x)$ & ${\cal O}_{L_1}$ & $\epsilon_2(B_x)$ & ${\cal O}_{L_2}$
                & $\epsilon_1(B_x)$ & ${\cal O}_{L_1}$ & $\epsilon_2(B_x)$ & ${\cal O}_{L_2}$\\ 
 \noalign{\smallskip} \hline\noalign{\smallskip}\hline
WENO+3
 &  32 &       2.49E-03 &   -    &       1.94E-04 &   -    &       7.81E-04 &   -    &       1.74E-05 &   -    \\ \noalign{\smallskip} 
 &  64 &       4.13E-04 &    2.6 &       1.73E-05 &    3.5 &       1.29E-04 &    2.6 &       1.12E-06 &    4.0 \\ \noalign{\smallskip} 
 & 128 &       5.72E-05 &    2.9 &       1.16E-06 &    3.9 &       1.82E-05 &    2.8 &       5.38E-08 &    4.4 \\ \noalign{\smallskip} 
 & 256 &       7.69E-06 &    2.9 &       7.24E-08 &    4.0 &           -    &   -    &            -   &   -    \\ \noalign{\smallskip} 
\hline
\Lim
 &  32 &       2.49E-03 &   -    &       1.94E-04 &   -    &       7.81E-04 &   -    &       1.74E-05 &   -    \\ \noalign{\smallskip} 
 &  64 &       4.13E-04 &    2.6 &       1.73E-05 &    3.5 &       1.29E-04 &    2.6 &       1.12E-06 &    4.0 \\ \noalign{\smallskip} 
 & 128 &       5.72E-05 &    2.9 &       1.16E-06 &    3.9 &       1.82E-05 &    2.8 &       5.38E-08 &    4.4 \\ \noalign{\smallskip} 
 & 256 &       7.69E-06 &    2.9 &       7.24E-08 &    4.0 &           -    &   -    &            -   &   -    \\ \noalign{\smallskip} 
\hline
WENO-Z
 &  32 &       8.17E-04 &   -    &       1.02E-04 &   -    &       1.63E-04 &   -    &       7.39E-06 &   -    \\ \noalign{\smallskip} 
 &  64 &       5.10E-05 &    4.0 &       2.89E-06 &    5.1 &       1.07E-05 &    3.9 &       1.50E-07 &    5.6 \\ \noalign{\smallskip} 
 & 128 &       1.83E-06 &    4.8 &       5.23E-08 &    5.8 &       3.78E-07 &    4.8 &       1.87E-09 &    6.3 \\ \noalign{\smallskip} 
 & 256 &       5.94E-08 &    4.9 &       8.28E-10 &    6.0 &           -    &   -    &           -    &   -    \\ \noalign{\smallskip} 
\hline
MP5
 &  32 &       9.57E-04 &   -    &       1.04E-04 &   -    &       1.96E-04 &   -    &       7.34E-06 &   -    \\ \noalign{\smallskip} 
 &  64 &       5.16E-05 &    4.2 &       3.02E-06 &    5.1 &       1.07E-05 &    4.2 &       1.53E-07 &    5.6 \\ \noalign{\smallskip} 
 & 128 &       1.75E-06 &    4.9 &       5.15E-08 &    5.9 &       3.66E-07 &    4.9 &       1.85E-09 &    6.4 \\ \noalign{\smallskip} 
 & 256 &       5.69E-08 &    4.9 &       8.04E-10 &    6.0 &           -    &   -    &           -    &   -    \\ \noalign{\smallskip} 
\hline
\end{tabular*}
\end{table}

%%%%%%%%%%%%%%%%%%%%%%%%%%%%%%%%%%%%%%%%%%%%%%%%%%%%%%%%%%%%%%%%
\subsubsection{Three Dimensional Propagation}
%
%%%%%%%%%%%%%%%%%%%%%%%%%%%%%%%%%%%%%%%%%%%%%%%%%%%%%%%%%%%%%%%%

We propose a novel three dimensional extension of the vortex 
problem, consisting of similar initial conditions as the 2D case, 
albeit the radius $r$ now refers to the spherical one, $r = \sqrt{x^2+y^2+z^2}$.
The perturbation of pressure is now given by 
\begin{equation}
 p = 1 + \frac{1}{4\,q}\,\Big[\mu^2\,\left(1 - 2\,q\,(r^2 - z^2)\right) - 
         \kappa^2\,\rho\Big]\,{\rm e}^{2\,q\,(1-r^2)}\,, 
\end{equation}
while we prescribe also a vertical velocity $v_z = 2$. 
The computational domain is the cube $[-5,5]^3$ 
with periodic boundary conditions. The evolution stops after 10 time units. 

The last four columns of Table \ref{table:L1_vortex} report the $L_1$ and 
$L_2$ norm errors of $B_x$ showing an excellent agreement with the analytical solution.
Notice that the errors measured in $L_2$ norm are systematically smaller than $L_1$
errors and a comparison between similar configurations using different norms
(as reported in \cite{DBFM08}) may be deceitful.
Keeping that in mind and given the somewhat diverse configurations, 
one can see that our results (in $L_2$ norm) are competitive with those of \cite{DBFM08}
at least at a qualitative level.

Divergence errors, shown in Fig \ref{fig:vortex}, quickly decrease as the mesh
thickens and fall below $10^{-8}$ at the resolution of $128^3$ for the fifth-order
schemes.

The computational cost is in accordance with previous tests,
giving a ratio of $1:0.99:1.46:1.24$ for WENO+3, \Lim, WENO-Z and MP5, 
respectively.

%%%%%%%%%%%%%%%%%%%%%%%%%%%%%%%%%%%%%%%%%%%%%%%%%%%%%%%%
\subsection{Advection of a magnetic field loop}
%
%
%%%%%%%%%%%%%%%%%%%%%%%%%%%%%%%%%%%%%%%%%%%%%%%%%%%%%%%%

We now consider the advection of a magnetic field loop. 
For sufficiently large plasma $\beta$, specifying a thermal 
pressure dominance, the loop is transported as a passive scalar. 
The preservation of the initial circular shape tests 
the scheme's dissipative properties and the correct 
discretization balance of multidimensional terms
\cite{GS05, GS08, LD09, MT10}.

%%%%%%%%%%%%%%%%%%%%%%%%%%%%%%%%%%%%%%%%%%%%%%%%%%%%%%%%%%%%%%%%
\subsubsection{Two Dimensional Propagation}
%
%%%%%%%%%%%%%%%%%%%%%%%%%%%%%%%%%%%%%%%%%%%%%%%%%%%%%%%%%%%%%%%%

Following \cite{GS05, FHT06}, the computational box is defined by $x\in[-1,1]$ 
and $y\in[-0.5,0.5]$ discretized on $2N_y\times N_y$ grid cells ($N_y=64$).  
Density and pressure are initially constant and equal to $1$. 
The velocity of the flow is given by 
$\vec{v} = V_0(\cos\alpha, \sin\alpha)$
with $V_0 = \sqrt{5}$, $\sin \alpha = 1/\sqrt{5}$ and
$\cos \alpha = 2/\sqrt{5}$. The magnetic field is defined through its 
magnetic vector potential as 
\begin{equation}  
A_z = \left\{ \begin{array}{ll}
    a_0 + a_2r^2 & \textrm{if} \quad 0 \leq r \leq R_1 \,, \\ \noalign{\medskip}  
    A_0(R-r) & \textrm{if} \quad R_1 < r \leq R \,, \\ \noalign{\medskip}
    0   & \textrm{if} \quad r > R \,,
  \end{array} \right.
\end{equation}     
where $A_0 = 10^{-3}$, $R=0.3$, $R_1=0.2 R$, $a_2 = -0.5A_0/R_1$, 
$a_0 = A_0(R-R_1) - a_2R_1^2$ and $r=\sqrt{x^2 + y^2}$. 
The modification to the vector potential in the $r \leq R_1$ region (with respect to 
similar setups presented by other investigators) is done to remove the singularity in 
the loop's center that can cause spurious oscillations and erroneous evaluations of the
magnetic energy. 
The simulations are allowed to evolve until $t=2$ ensuring the
crossing of the loop twice through the periodic boundaries. 

\begin{figure}\centering
\includegraphics[width=0.9\textwidth]{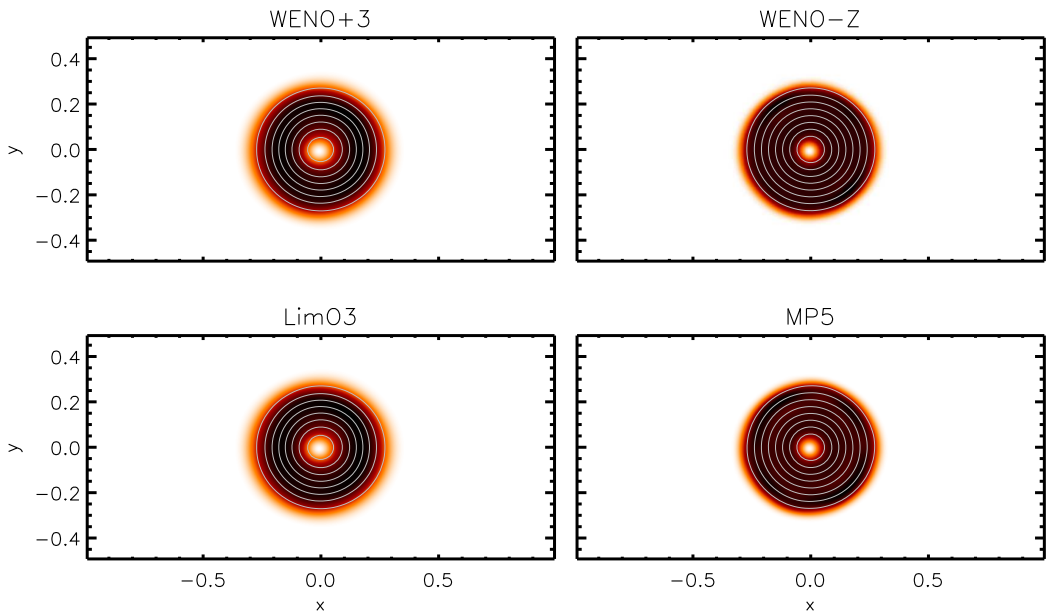}
 \caption{Magnetic energy density for the 2D field loop problem at $t=2$
  computed with the third-order (left) and fifth-order (right) schemes 
  at the resolution of $128\times 64$ points with Courant number $C_a=0.4$. 
  Magnetic field lines are overplotted using 9 contour levels equally 
  spaced between $10^{-5}$ and $10^{-3}$.}
 \label{fig:fl2d}
\end{figure}

\begin{figure}\centering
\includegraphics[width=0.45\textwidth]{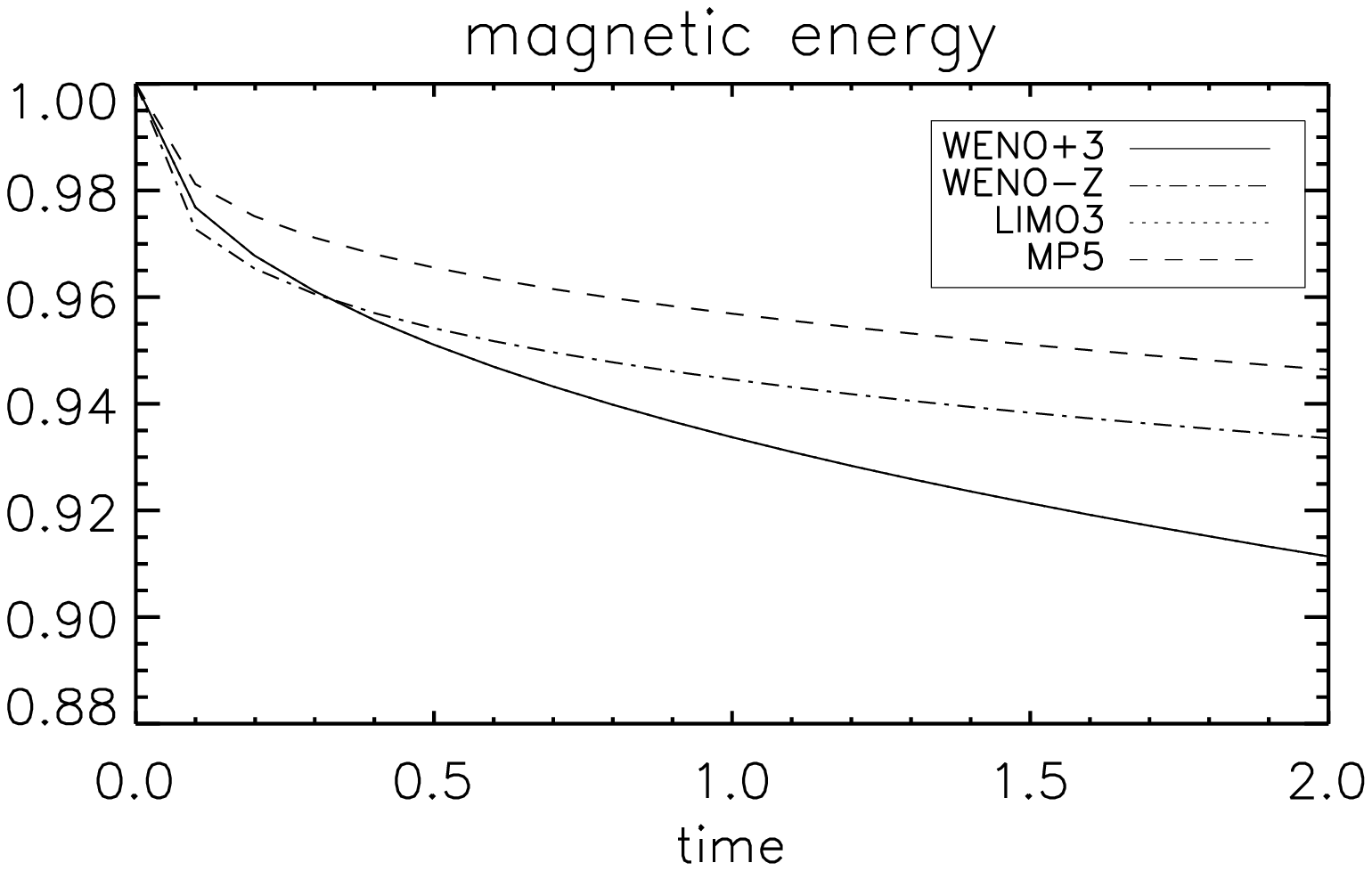}
\includegraphics[width=0.45\textwidth]{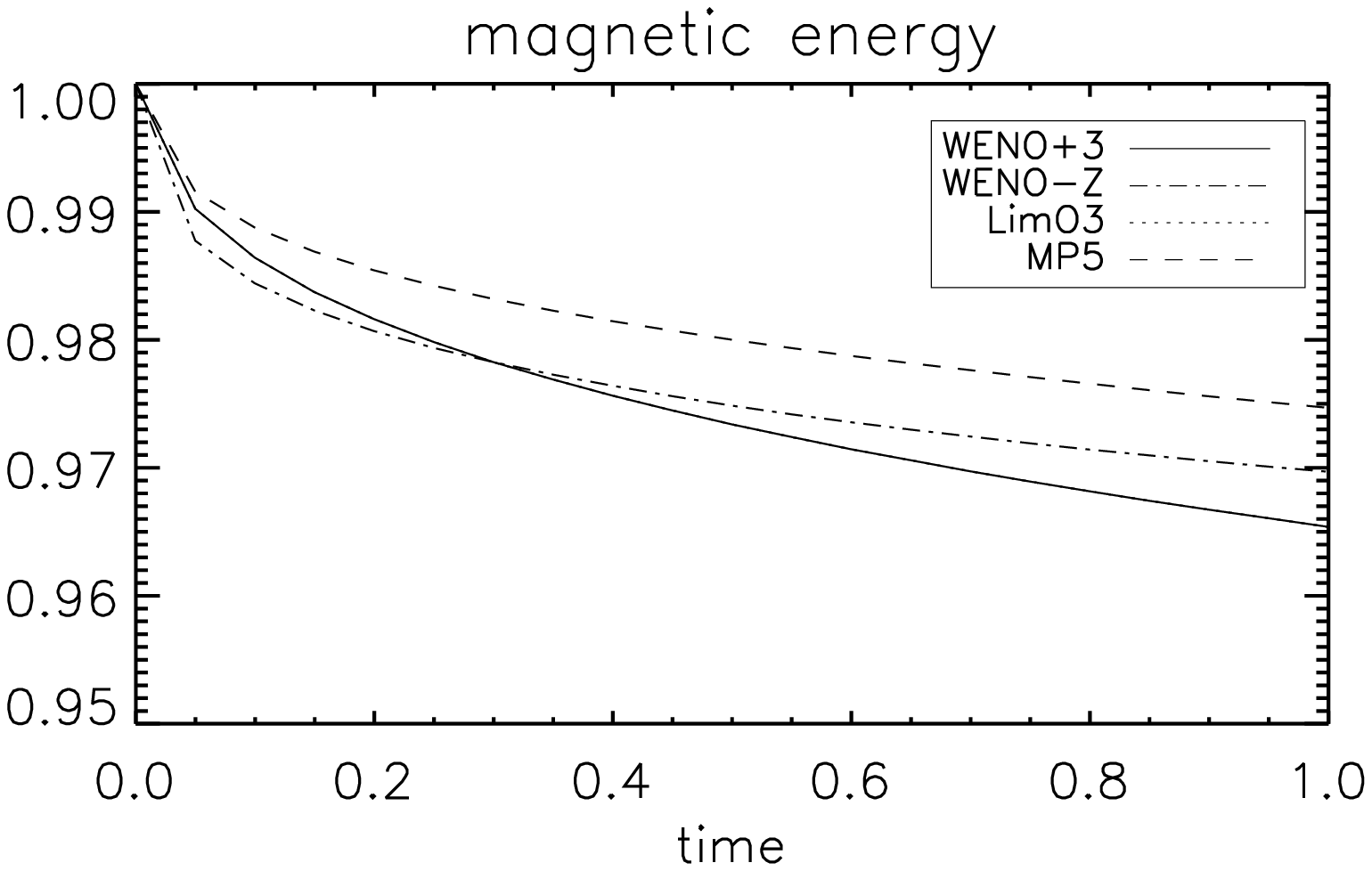}
 \caption{Time evolution of the magnetic energy density, normalized to its 
  initial value, for the 2D (left) and 3D (right) field loop problem
  at the resolution of $128\times 64$ grid points. 
  The magnetic energy is better conserved for the MP5 method. \Lim and 
  WENO+3 show no pronounced difference for this particular problem.}
 \label{fig:fl2d_energy}
\end{figure}

In Fig. \ref{fig:fl2d} the magnetic energy density is displayed
for the \Lim, WENO+3, WENO-Z and MP5 schemes, along with iso-contours 
of the $z$ component of the magnetic vector potential. 
The initial circular shape is preserved well by all schemes. The third-order 
schemes are substantially more diffusive, as can be seen on the borders of the loop. 
This is confirmed by the time evolution of the magnetic energy density (normalized 
to its initial value), plotted in the left panel of Fig.\ref{fig:fl2d_energy}. 
The power law behaviour is similar for the schemes of the same order, with the MP5
method being the least diffusive. No pronounced difference is found between the \Lim
and WENO+3 schemes, for this particular problem. 

\begin{figure}\centering
\includegraphics[width=\textwidth]{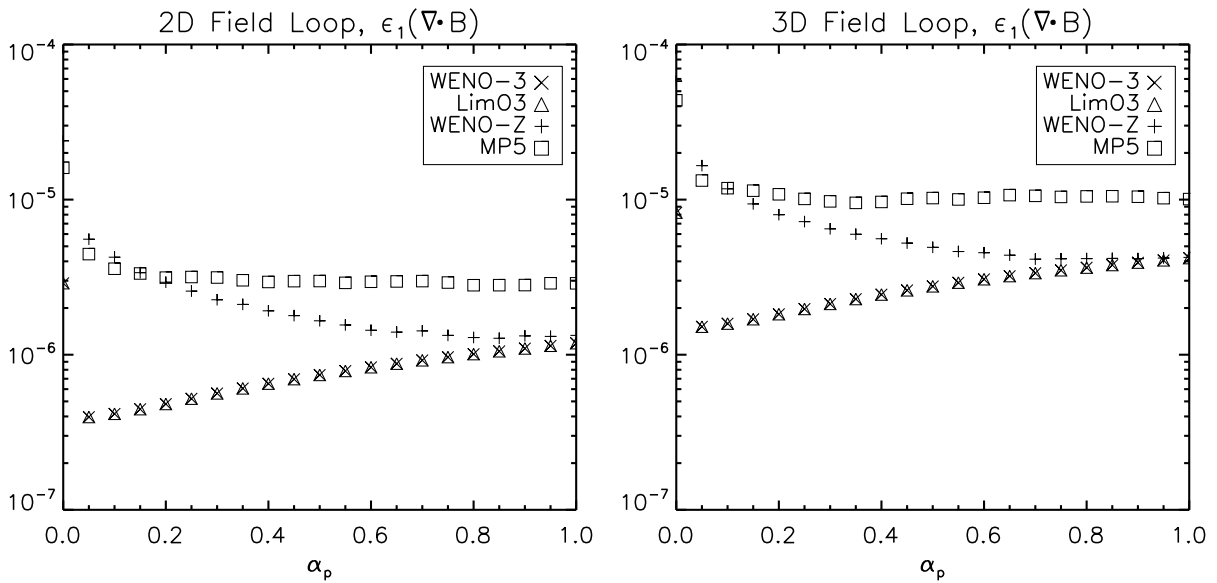}
 \caption{Divergence errors as function of the damping parameter $\alpha_p$ 
          for the field loop test problem in 2D (left, $t=2$) on $128\times 64$ 
          grid points and 3D (right, $t=1$) on $128\times 128\times 256$ grid
          points.}
 \label{fig:fl_alpha}
\end{figure}

The divergence of magnetic field measured in $L_1$ norm
is shown in the left panel of Fig \ref{fig:fl_alpha}, as a function of 
$\alpha_p\in[0,1.0]$. For the fifth-order schemes errors are minimized 
when $\alpha_p\gtrsim 0.4$ whereas \Lim and WENO+3 present 
smaller errors for $\alpha_p\lesssim 0.2$.

%%%%%%%%%%%%%%%%%%%%%%%%%%%%%%%%%%%%%%%%%%%%%%%%%%%%%%%%%%%%%%%%
\subsubsection{Three Dimensional Propagation}
%
%%%%%%%%%%%%%%%%%%%%%%%%%%%%%%%%%%%%%%%%%%%%%%%%%%%%%%%%%%%%%%%%

The three dimensional version of this problem is particularly challenging as the 
correct evolution depends on how accurately the $\nabla\cdot\vec{B}=0$ condition 
is preserved and how the multidimensional MHD terms are balanced out.
The computational domain $-0.5\le x \le 0.5$, $-0.5\le y \le 0.5$,
$ -1.0\le z \le 1.0$ is resolved onto $128\times 128\times 256 $ 
zones. As for the two-dimensional case the vector potential $A_3$ is used
to initialize the magnetic field,  which is then rotated 
using the coordinate transformation given by Eq. (\ref{eq:rot_mat}) with
$\alpha = 0$ and $\gamma = \tan^{-1}1/2$. Even though the loop is rotated only 
around one axis, the velocity profile $(v_x,v_y,v_z)=(1,1,2)$ makes the test 
intrinsically three-dimensional.
Once again, pressure and density are taken uniform and equal to unity while
boundary conditions are periodic in all directions.

\begin{figure}\centering
\includegraphics[width=\textwidth]{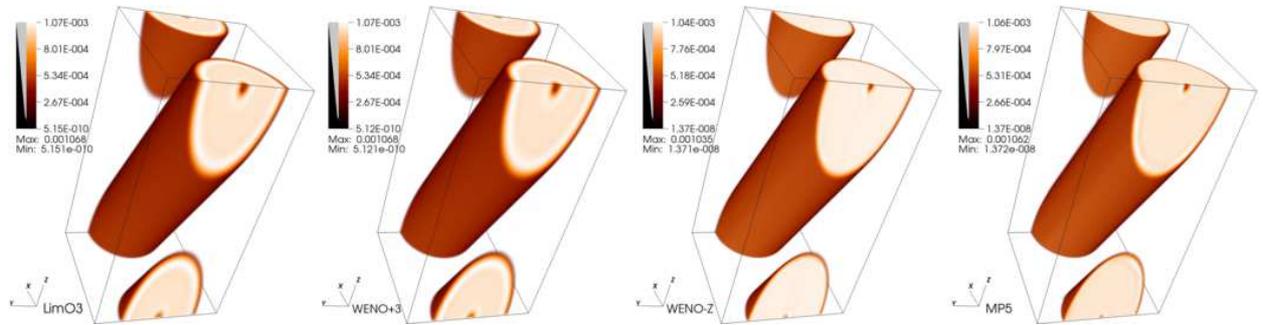}
 \caption{Magnetic energy density for the 3D field loop problem at $t=1$
  computed on $128\times 128\times 256$ grid zones with Courant number 
  $0.3$.
  From left to right: \Lim, WENO+3, WENO-Z, MP5.
  All schemes preserve the circularity of the loop, with the fifth-order 
  schemes displaying sharper borders.}
 \label{fig:fl3d}
\end{figure}

The preservation of the loop's shape can be seen in Fig. \ref{fig:fl3d}. All
schemes preserve the shape, with \Lim and WENO+3 being equally more diffusive 
(notice the thickness of the dark area at the loop's borders, as well as the 
brighter ring just inside the loop).
As for the 2D case, one can see that MP5 is the least diffusive in preserving 
the magnetic energy (right panel of Fig. \ref{fig:fl2d_energy}), while 
the dissipation rates for \Lim and WENO+3 practically coincide. 
Moreover, the three-dimensional $L_1$ norm error of $\nabla\cdot\vec{B}$ 
(right panel of Fig \ref{fig:fl_alpha}) exhibits a behaviour similar to 
the two dimensional case.
As before, the relative CPU scaling between WENO+3, \Lim, WENO-Z and MP5 for
this test problem is $1:0.97:1.45:1.25$. 

%%%%%%%%%%%%%%%%%%%%%%%%%%%%%%%%%%%%%%%%%%%%%%%%
\subsection{Orszag-Tang}
%
%
%%%%%%%%%%%%%%%%%%%%%%%%%%%%%%%%%%%%%%%%%%%%%%%%

%
%    Paper          going until t=1    Showing pr cut   Resolution
%
%  Balsara98a           YES (Iso)        -
%  BTW04                -                YES            384 x 384
%  Dumbser              YES              NO             200 x 200  on [0, 2\pi]
%  Jiang Wu 99          -                YES            192 x 192  on [0, 2\pi]
%  Lee Deane 09         YES               -             400 x 400  on [0,1]
%  Li08                 -                YES            256 x 256  on [0, 2\pi]
%  LdZ00                -                YES            192 x 192  on [0, 1]
%  Mignone07            Yes (Iso)         -             400 x 400  
%  MK05                 -                YES
%  Rossmanith           -                YES            256 x 256  on [0, 2\pi]
%  Stone(Athena)        -                YES            192 x 192  on [0, 1]
%  ZYL06                -                YES            193 x 193

\begin{figure}[!h]
\centering
\includegraphics[width=\textwidth]{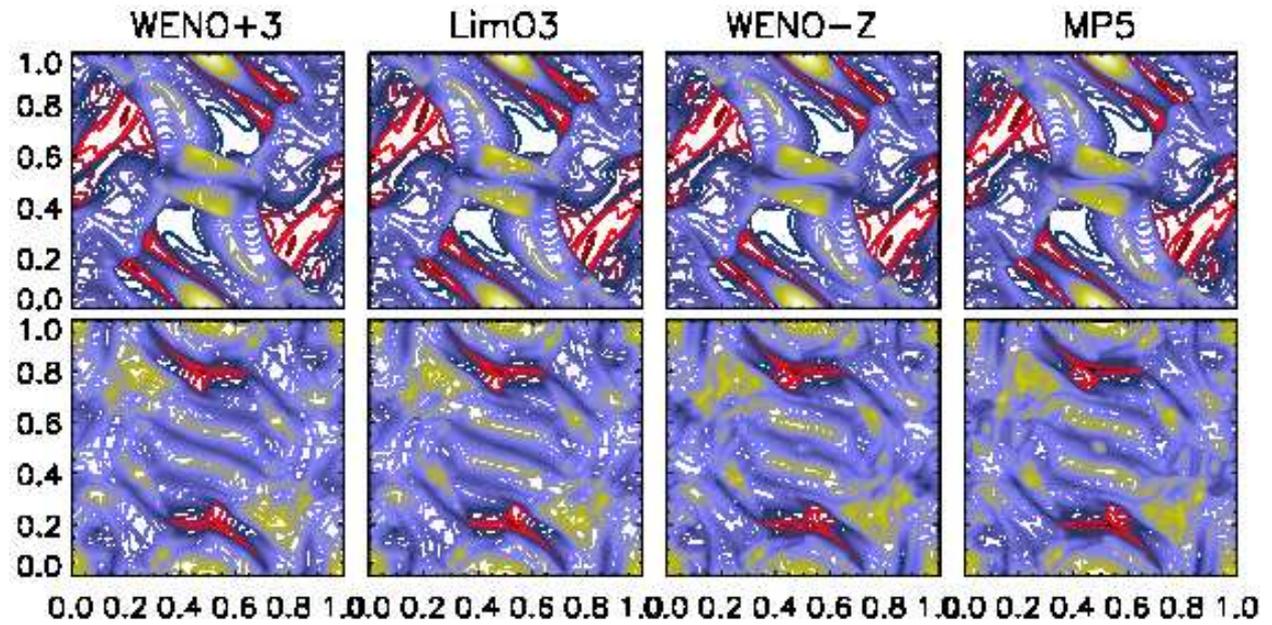}
 \caption{Density contour plots for the Orszag-Tang system at 
          $t=0.5$ (top) and $t=1$ (bottom) for the 
          selected schemes using $256^2$ grid points. 
          Thirty equally spaced levels ranging from 
          $0.3831 \Gamma^2$ to $2.2414\Gamma^2$ for the top panel
          and from $0.1944\Gamma^2$ to $1.9337\Gamma^2$ for the
          bottom panel are shown.}
 \label{fig:ot}
\end{figure}

\begin{figure}[!h]
\centering
\includegraphics[width=\textwidth]{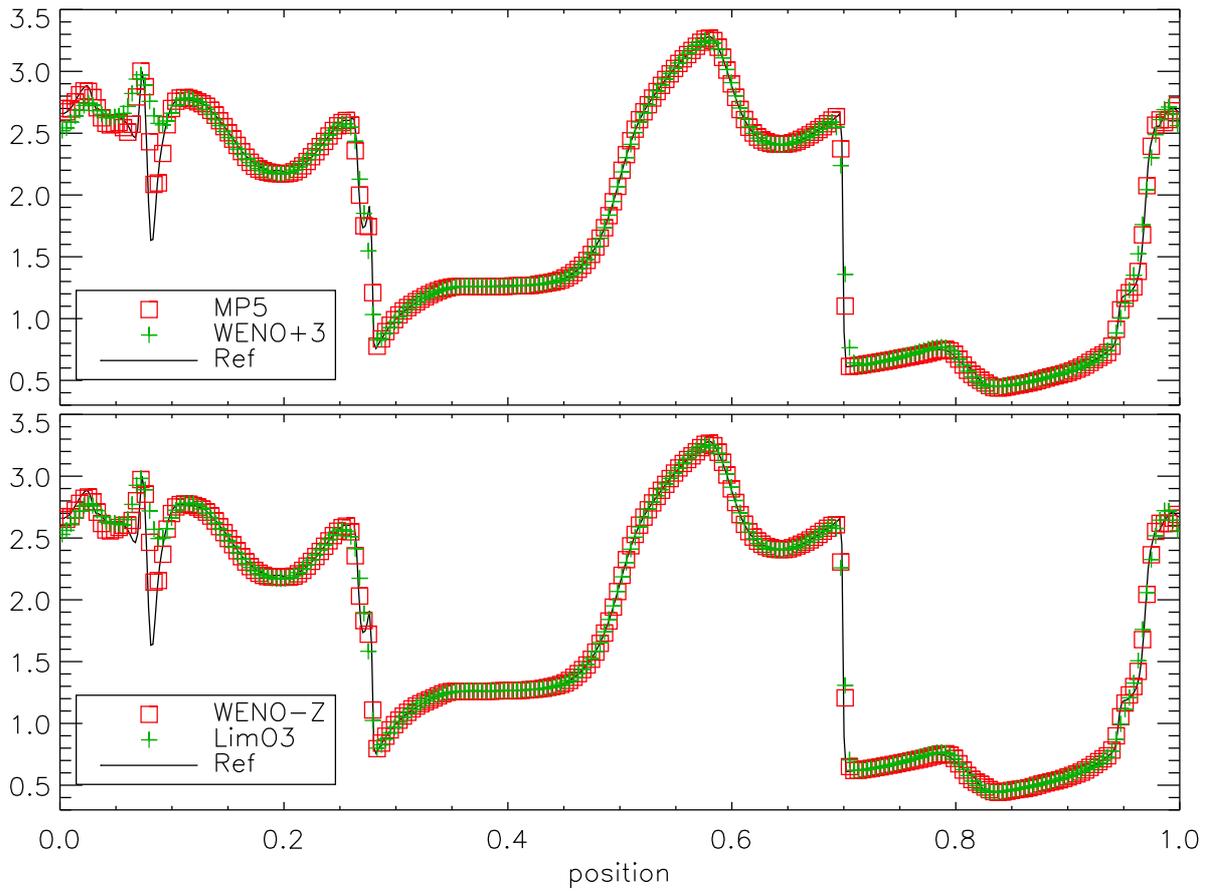}
 \caption{Horizontal cut at $y=0.3125$ showing gas pressure in 
          the Orszag-Tang system at $t=0.5$ at the resolution of
          $256^2$. MP5 and WENO+3
          are shown in top panel (squares and plus signs),
          WENO-Z and \Lim in the bottom. The solid line gives 
          a reference solution obtained with second-order 
          constrained transport algorithm on $1024^2$ zones.}
 \label{fig:ot_slice}
\end{figure}

The Orszag-Tang vortex system describes a doubly periodic 
fluid configuration leading to two-dimensional supersonic MHD turbulence.
The domain $[0,1]^2$ is initially filled with constant density and pressure
respectively equal to $\rho=\Gamma^2$ and $p=\Gamma$, while
velocity and magnetic field are initialized to 
$\vec{v}=\left(-\sin 2\pi y, \sin2\pi x, 0\right)$ and
$\vec{B}=\left(-\sin 2\pi y, \sin4\pi x, 0\right)$, respectively.
Although an analytical solution is not known, its simple and reproducible 
set of initial conditions has made it a widespread benchmark for 
inter-scheme comparison, see for example \cite{Toth00}.
Density contour plots, as in \cite{LD09} are shown in the top and bottom rows 
of Fig. \ref{fig:ot} at $t=0.5$ and $t=1$, respectively, using 
a resolution of $256^2$ points.
The dynamics is regulated by multiple shock interactions leading to the 
formation of small scale vortices and density fluctuations.
Our results at $t=0.5$ are in good agreement with previous investigations, e.g. 
\cite{JW99, Toth00, LdZ04, Ross06, LD09}, with WENO+3 and \Lim  
showing increased numerical dissipation when compared to WENO-Z and MP5. 
This is further confirmed in Fig \ref{fig:ot_slice} where horizontal cuts 
at $y=0.3125$ in the pressure distribution are plotted against a reference 
solution obtained with the second-order CT-CTU scheme of \cite{GS05} 
on a finer mesh ($1024^2$), see also \cite{JW99, LdZ00, Ross06}. 

The most noticeable difference occurs at $t=1$, when the fifth-order schemes 
(in particular, MP5) reveal the formation of a central magnetic island featuring 
a high density spot also recognizable in the results of \cite{LD09} and in 
\cite{B98,M07} for the isothermal case.
This structure is absent in the third-order schemes and may be induced 
by the decreased effective resistivity across the central current sheet, 
as discussed in \cite{M07}.

\begin{figure}[!h]
\centering
\includegraphics[width=\textwidth]{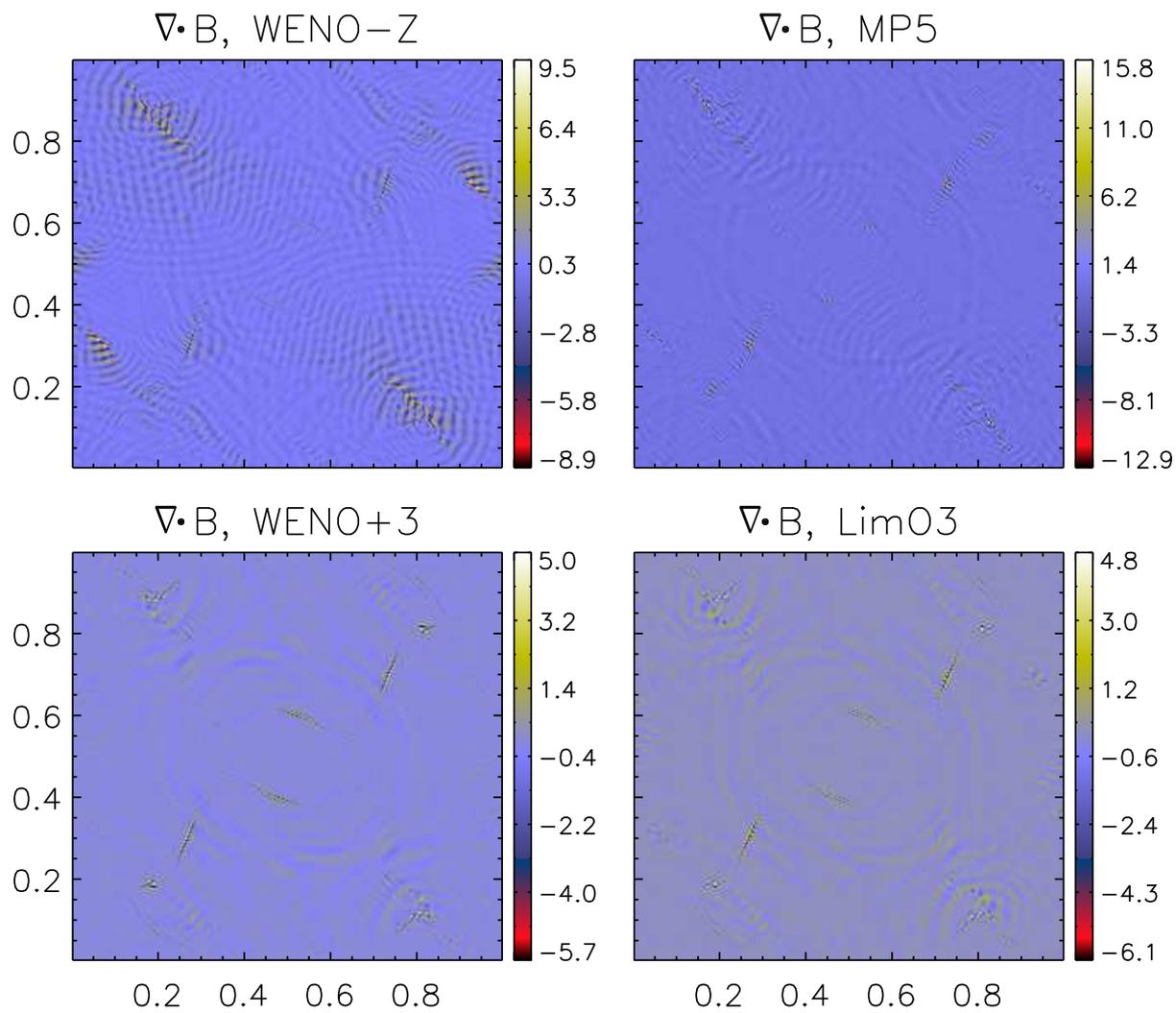}
 \caption{Divergence errors for the four selected scheme at $t=0.5$ 
          on $256^2$ grid zones.}
 \label{fig:ot_divb}
\end{figure}

Divergence errors, shown in Fig \ref{fig:ot_divb} at $t=0.5$,
are comparable with those given by other investigators (e.g.
\cite{Ross06, Li08}) and reach their maximum magnitude in presence
of discontinuous features.

The computational cost of \Lim, WENO-Z and MP5 relative 
to that of WENO+3 ($=1$) are found to be $1.01:1.47:1.29$, in
analogy with the previous results.

%%%%%%%%%%%%%%%%%%%%%%%%%%%%%%%%%%%%%%%%%%%%%%%%
\subsection{Kelvin-Helmholtz Unstable Flows}
%
%
%%%%%%%%%%%%%%%%%%%%%%%%%%%%%%%%%%%%%%%%%%%%%%%%

As a final example, we propose the nonlinear evolution of
the Kelvin-Helmholtz instability in two dimensions.
The base flow consists of a single shear layer with an initially 
uniform magnetic field lying in the $xz$ plane at
an angle $\theta=\pi/3$ with the direction of propagation:
\begin{equation}\label{eq:KH}
 \vec{v} = \left[\frac{M}{2}\tanh\left(\frac{y}{y_0}\right),\,0,\,0\right] 
              \,,\quad
 \vec{B} = c_a\sqrt{\rho}\Big[\cos\theta,\, 0,\, \sin\theta\Big]\,,
\end{equation}
where $M=1$ is the Mach number, $y_0=1/20$ is the steepness of the shear, 
$c_a = 0.1$ is the Alfv\'en speed.
Density and pressure are initially constant and equal to $\rho = 1$ and $p=1/\Gamma$.
A single-mode perturbation 
$v_y= v_{y0}\sin\left(2\pi x\right)\exp\left[-y^2/\sigma^2\right]$
with $v_{y0} = 10^{-2}$, $\sigma = 0.1$ is super-imposed
as in \cite{MBR96}.
Computations are carried out in a Cartesian box $[0,1]\times[-1,1]$
for $t=20$ time units on a $N_x\times 2N_x$ mesh, where $N_x=64,128,256$.

\begin{figure}[!h]
\centering
\includegraphics[width=0.8\textwidth]{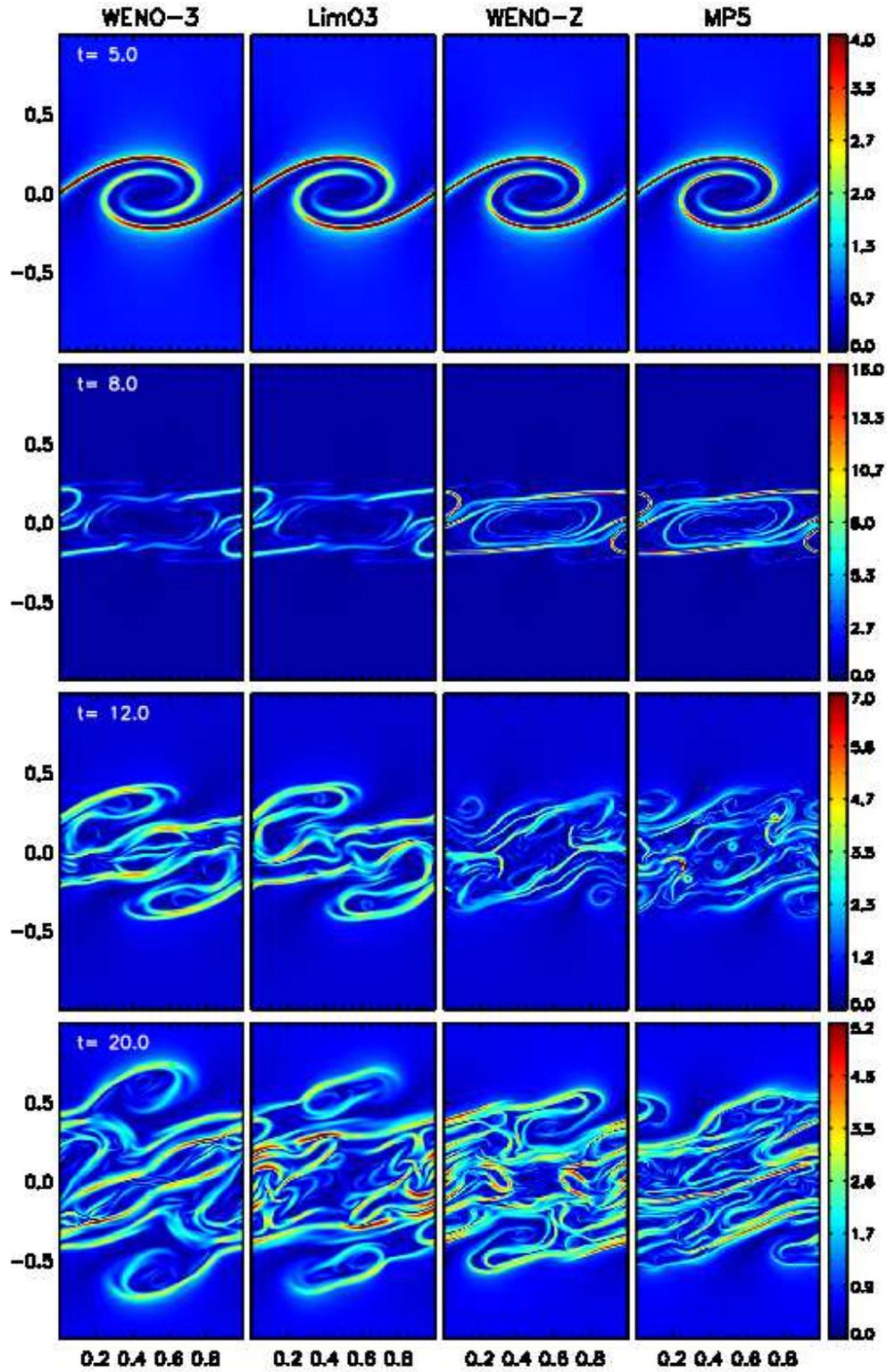}
 \caption{Snapshots of the evolution of the Kelvin-Hemlholtz unstable
          layer at $t=5$ (first panel from top), $t=8$ (second panel),
          $t=12$ (third panel) and $t=20$ (bottom panel).
          The images show the ratio of the poloidal field strength and
          the toroidal component, $\sqrt{B_x^2+B_y^2}/B_z$.
          Left to right columns corresponds to computations obtained with
          WENO+3, \Lim, WENO-Z and MP5, respectively, at the resolution 
          of $256\times 512$. Note how the colorbar maximum value changes 
          at different instant to reflect the corresponding magnetic field 
          strength.}
 \label{fig:khsnap}
\end{figure}

\begin{figure}
\centering
\includegraphics[width=0.8\textwidth]{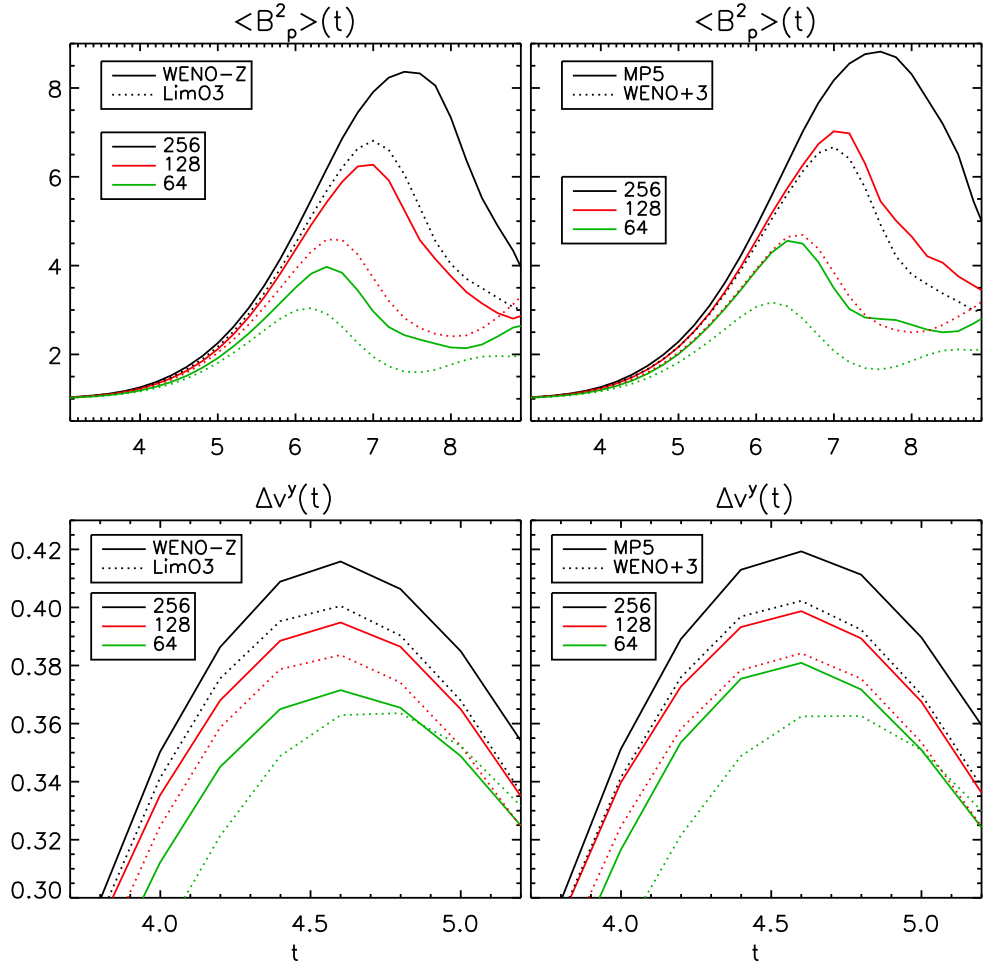}
 \caption{Volume integrated magnetic energy (top panels) and 
          growth rate (computed as $\Delta v^y=(v^y_{\max}-v^y_{\min})/2$)
          as functions of time.
          Here $B^2_p=B^2_x+B^2_y$ accounts for the 
          "poloidal" contribution only. Solid and dotted lines
          corresponds to integrations carried with WENO-Z 
          and \Lim (left panels), MP5 and WENO+3 (right panels).
          The different colors, green, red and black indicate 
          different numerical resolution, i.e., $64$, $128$ and $256$,
          respectively. }
 \label{fig:khgrowth}
\end{figure}

The evolutionary stages are shown in Fig \ref{fig:khsnap}, where
we display color maps of the ratio $(B_x^2+B_y^2)^{\HALF}/B_z$ at the 
largest resolution $256\times 512$ for WENO+3, \Lim, WENO-Z and MP5.
For $t\lesssim 5$ the perturbation follows a linear growth phase
during which magnetic field lines wound up through the formation 
of a typical cat's eye vortex structure, \cite{MBR96, JGRF97}, 
see the top row in Fig \ref{fig:khsnap}.
During this phase, magnetic field lines become distorted 
all the way down to the smaller diffusive scales and 
the resulting field amplification becomes larger for 
higher magnetic Reynolds numbers.
As such, we observe in the top row of Fig \ref{fig:khgrowth} that the 
magnetic energy grows faster not only as the resolution is increased
from $64$ to $256$ mesh points (green, red, black), 
but also when switching from a third-order to a fifth order scheme
(solid vs. dotted lines).
In particular, one can see that half of the grid resolution is 
needed by MP5 to match the results obtained with WENO+3.
A somewhat lesser gain can be inferred by comparing WENO-Z and
\Lim.
Similarly, the growth rate (computed as 
$\Delta v^y= (v^y_{\max} - v^y_{\min})/2$
see bottom panel in Fig. \ref{fig:khgrowth}), is closely
related to the poloidal field amplification and evolves
faster for smaller numerical resistivity and thus for finer grids
and/or less dissipative schemes. 

Field amplification is eventually prevented when $t\gtrsim 8$ by 
tearing mode instabilities leading to reconnection events capable 
of expelling magnetic flux from the vortex 
(second row in Fig. \ref{fig:khsnap}), \cite{JGRF97}.
Throughout the saturation phase (third and fourth row in Fig 
\ref{fig:khsnap}) the mixing layer enlarges and the
field lines thicken into filamentary structures. 
During this phase one can clearly recognize that small scale 
structures are best spotted with the fifth-order methods
while they appear to be more diffused with WENO+3 and \Lim.

The CPU costs relative to that of WENO+3 ($=1$) follow the ratios 
$0.98:1.48:1.24$ for \Lim, WENO-Z and MP5, respectively, and confirm the 
same trend already established in previous tests.

%%%%%%%%%%%%%%%%%%%%%%%%%%%%%%%%%%%%%%%%%%%%%%%%%%%%%%%%%%%%%
\section{Conclusions}
\label{sec:conclusions}
%
%
%
%%%%%%%%%%%%%%%%%%%%%%%%%%%%%%%%%%%%%%%%%%%%%%%%%%%%%%%%%%%%%%

We have presented a class of high-order finite difference schemes for 
the solution of the compressible ideal MHD equations in multiple 
spatial dimensions.
The numerical framework adopts a point-wise, cell centered 
representation of the primary flow variables and 
has been conveniently cast in conservation 
form by providing highly accurate interface values through a 
one-dimensional finite volume reconstruction approach. 
The divergence-free condition of magnetic field is monitored by 
introducing a scalar generalized Lagrange multiplier, as in \cite{Dedner02}, offering 
propagation as well as damping of divergence errors in a mixed 
hyperbolic/parabolic way.
This greatly simplifies the task of obtaining highly accurate solutions 
since the reconstruction process can be carried out on one-dimensional 
stencils using the information available at cell centers. 
In this respect, our formulation completely avoids expensive elliptic 
cleaning steps, does not require genuinely multidimensional 
interpolation and eludes the complexities required by staggered 
mesh algorithms.
Selected numerical schemes based on third- as well as 
fifth-order accurate constraints have been presented and compared.

\begin{itemize}

\item 
The recently improved version of the third-order WENO scheme (WENO+3, 
\cite{YC09}) and the \Lim reconstruction based on new limiter 
functions (introduced in \cite{CT09}) perform equally well exhibiting 
third-order accuracy in smooth problems and non-oscillatory 
transitions at discontinuities.

\item 
The new fifth-order WENO scheme (WENO-Z, see \cite{BCCS08}) and 
the monotonicity preserving algorithm (MP5) of \cite{SH97}
yield high-quality results on all of the selected tests and report
orders of accuracy close to $5$ for multidimensional smooth problems.
Both WENO-Z and MP5 perform with a greatly reduced amount 
of numerical dissipation and provide highly accurate solution
with much fewer grid points when compared to third-order accurate 
schemes.
Still, we have found MP5 to give slightly better results WENO-Z
in terms of reduced computational cost, improved accuracy and 
sharper transitions at discontinuous fronts.

\item
Fifth-order schemes are found to be $\lesssim 50$ (for WENO-Z) and 
$\lesssim 30$ (for MP5) per cent slower 
than third-order ones, depending on the particular choice.
This favorably advocates towards the use of higher order schemes rather
than lower order ones, since the same level of accuracy can be attained 
at a much lower resolution still giving a tremendous gain 
in computing time.
For three-dimensional problems, for example, the gain can be almost
two orders of magnitude in CPU cost. 

\item
The results obtained with the present finite difference
formulation are competitive (in terms of accuracy and 
description of discontinuities) with recently developed FV 
schemes (e.g., \cite{DBFM08, Balsara09, BRDM09})
and noticeably improve over traditional $2^{\rm nd}$ order 
Godunov-type schemes in terms of reduced numerical 
dissipation.
The benefits offered by a high-order method such as the
ones presented here are particularly relevant in the context
of MHD applications involving both smooth and discontinuous 
flows. 

\end{itemize}

\vspace*{2ex}\par\noindent
{\bf Acknowledgements.}

Extensive numerical testing of the finite difference
schemes presented in this paper was made possible by 
the computational facilities available thanks to the 
INAF-CINECA agreement.

%%%%%%%%%%%%%%%%%%%%%%%%%%%%%%%%%%%%%%%%%%%
\appendix
%
%
%%%%%%%%%%%%%%%%%%%%%%%%%%%%%%%%%%%%%%%%%%%
%%%%%%%%%%%%%%%%%%%%%%%%%%%%%%%%%%%%%%%%%%%%%%%%%%%%%%%%%%%%%%%%%%%
\section{Conservative eigenvectors of the GLM-MHD Equations}
\label{app:eigenv}
%
%
%
%
%%%%%%%%%%%%%%%%%%%%%%%%%%%%%%%%%%%%%%%%%%%%%%%%%%%%%%%%%%%%%%%%%%

The $9\times 9$ matrix  of the conservative MHD equations
in one dimension introduced can be decomposed, given the eigenvalues 
(see Eq. \ref{eq:eigenvalues}), to the corresponding left and 
right eigenvectors. Following partially the notation of \cite{RB96,JW99}, we define
\begin{equation}
\alpha_{f}^2 = \frac{a^2-c_{s}^2}{c_{f}^2-c_{s}^2}, \quad\quad \alpha_{s}^2 = \frac{c_{f}^2 - a^2}{c_{f}^2-c_{s}^2}, \quad\quad \beta_y = \frac{B_y}{\sqrt{B_{y}^2+B_{z}^2}}, \quad\quad \beta_z = \frac{B_z}{\sqrt{B_{y}^2+B_{z}^2}} 
\end{equation}
where $a =\sqrt{\Gamma p/\rho}$ denotes the speed of sound. With this notation,
the right eigenvectors in matrix form will be given by
\begin{equation}
  \tens{R} = \left(\begin{array}{ccccccccc}
   0  &          \alpha_f                                     &              0                 &        \alpha_s                 & 1              & \alpha_s                        &             0                  &  \alpha_f                          & 0   \\ \noalign{\medskip}
   0  &      \alpha_f \lambda_2                                 &              0                 &      \alpha_s \lambda_4           & v_x            & \alpha_s \lambda_6                &             0                  &  \alpha_f \lambda_8                 & 0   \\ \noalign{\medskip}
   0  &      \alpha_f v_y + J_{f0} \beta_y                    & - \beta_z S                    &  \alpha_s v_y - J_{s0} \beta_y  & v_y            & \alpha_s v_y + J_{s0} \beta_y   & - \beta_z S                    &  \alpha_f v_y - J_{f0} \beta_y     & 0   \\ \noalign{\medskip}
   0  &      \alpha_f v_z + J_{f0} \beta_z                    &   \beta_y S                    &  \alpha_s v_z - J_{s0} \beta_z  & v_z            & \alpha_s v_z + J_{s0} \beta_z   &   \beta_y S                    &  \alpha_f v_z - J_{f0} \beta_z     & 0   \\ \noalign{\medskip}
   1  &                0                                      &              0                 &             0                   & 0              &             0                   &             0                  &           0                        & 1   \\ \noalign{\medskip}
   0  &         J_{f1}\beta_y                                 &  - \beta_z \rho^{-\frac{1}{2}} &- J_{s1} \beta_y                 & 0              & - J_{s1} \beta_y                &   \beta_z \rho^{-\frac{1}{2}}  &  J_{f1} \beta_y                    & 0   \\ \noalign{\medskip}
   0  &         J_{f1}\beta_z                                 &    \beta_y \rho^{-\frac{1}{2}} &- J_{s1} \beta_z                 & 0              & - J_{s1} \beta_z                & - \beta_y \rho^{-\frac{1}{2}}  &  J_{f1} \beta_z                    & 0   \\ \noalign{\medskip}
   0  &               H_f - \Gamma_f                          &     - \Gamma_a                 &              H_s  - \Gamma_s    &\frac{1}{2} v^2 &        H_s + \Gamma_s           &     - \Gamma_a                 &    H_f + \Gamma_f                  & 0   \\ \noalign{\medskip}
 -c_h &                0                                      &              0                 &                        0        & 0              &             0                   &             0                  &              0                     & c_h \\ \noalign{\medskip}
\end{array}\right)\, 
\end{equation}
where $S=\rm{sign}(B_x)$, $H_{f,s} = \alpha_{f,s}(0.5v^2 + c_{f,s}^2 - \gamma_2 a^2)$, 
$J_{f,s0} = \alpha_{s,f} c_{s,f} S$ and $J_{f,s1} = \alpha_{s,f} a \rho^{\frac{1}{2}}$. 

On the other hand, the left eigenvectors are given by

\begin{equation}\label{eq:Left1}
  \tens{L}_{1,9} =
 \left( \begin{array}{c}
  0 \\ \noalign{\medskip}
  0 \\ \noalign{\medskip}
  0 \\ \noalign{\medskip}
  0 \\ \noalign{\medskip}
  \frac{1}{2}\\ \noalign{\medskip}
  0 \\ \noalign{\medskip}
  0 \\ \noalign{\medskip}
  0 \\ \noalign{\medskip}
  \mp \frac{1}{2c_h}\\ \noalign{\medskip}
\end{array}\right)^T
\,,\quad
\tens{L}_{3,7} =
 \left( \begin{array}{c}
  0.5 \Gamma_a \\ \noalign{\medskip}
  0 \\ \noalign{\medskip}
- 0.5 \beta_z S \\ \noalign{\medskip}
  0.5 \beta_y S \\ \noalign{\medskip}
  0 \\ \noalign{\medskip}
\mp 0.5 \sqrt{\rho} \beta_z \\ \noalign{\medskip}
\pm 0.5 \sqrt{\rho} \beta_y \\ \noalign{\medskip}
  0 \\ \noalign{\medskip}
  0 \\ \noalign{\medskip}
\end{array}\right)^T
\,,\quad
  \tens{L_{5}} =
 \left( \begin{array}{c}
 1 - 0.5 \tau v^2 \\ \noalign{\medskip}
 \tau v_x \\ \noalign{\medskip}
 \tau v_y \\ \noalign{\medskip}
 \tau v_z \\ \noalign{\medskip}
  0\\ \noalign{\medskip}
 \tau B_y \\ \noalign{\medskip}
 \tau B_z \\ \noalign{\medskip}
-\tau \\ \noalign{\medskip}
  0\\ \noalign{\medskip}
\end{array}\right)^T
\,, 
\end{equation}
\begin{equation}\label{eq:Left2}
  \tens{L}_{2,8} =
\frac{1}{2a^2} \left( \begin{array}{c}
  \gamma_1\alpha_f v^2 \pm \Gamma_f  \\ \noalign{\medskip}
  I_{fv_x} \mp \alpha_f c_f  \\ \noalign{\medskip}
  I_{fv_y} \pm J_{f0} \beta_y  \\ \noalign{\medskip}
  I_{fv_z} \pm J_{f0} \beta_z  \\ \noalign{\medskip}
  0\\ \noalign{\medskip}
  I_{fB_y} + J_{f1} \rho \beta_y  \\ \noalign{\medskip}
  I_{fB_z} + J_{f1} \rho \beta_z  \\ \noalign{\medskip}
  \alpha_f (\Gamma - 1)  \\ \noalign{\medskip}
  0 \\ \noalign{\medskip}
\end{array}\right)^T
\,,\quad
\tens{L}_{4,6} =
\frac{1}{2a^2} \left( \begin{array}{c}
  \gamma_1 \alpha_s v^2 \pm \Gamma_s  \\ \noalign{\medskip}
  I_{sv_x} \mp \alpha_s c_s  \\ \noalign{\medskip}
  I_{sv_y} \mp J_{s0} \beta_y  \\ \noalign{\medskip}
  I_{sv_z} \mp J_{s0} \beta_z  \\ \noalign{\medskip}
  0 \\ \noalign{\medskip}
  I_{sB_y} - J_{s1} \rho \beta_y  \\ \noalign{\medskip}
  I_{sB_z} - J_{s1} \rho \beta_z  \\ \noalign{\medskip}
  \alpha_s (\Gamma - 1)  \\ \noalign{\medskip}
  0 \\ \noalign{\medskip}
\end{array}\right)^T
\,. 
\end{equation}
where we prescribe $\tau = (\Gamma - 1)/a^2$, $\gamma_1  = (\Gamma - 1)/2$, $\gamma_2  = (\Gamma - 2)/(\Gamma - 1)$ and $I_{(f,s)\,(v_i,B_i)} = \Gamma^{-1} \alpha_{f,s} (v_i,B_i)$, with $i=x,y,z$.



\end{document}